\documentclass[preprint,12pt]{elsarticle}

\usepackage{amsmath,amssymb,bm}
\usepackage{xcolor}
\usepackage{booktabs}
\usepackage{array}
\usepackage{float}
\usepackage{algorithm}
\usepackage{algpseudocode}
\usepackage{tikz}
\usetikzlibrary{positioning,fit,arrows.meta,backgrounds}
\usepackage[colorlinks=true,linkcolor=blue,citecolor=blue,urlcolor=blue]{hyperref}

\tikzset{
  stage/.style={draw, rounded corners=2pt, align=center, inner sep=4pt, font=\small},
  cpu/.style={stage, fill=black!7},
  gpu/.style={stage, fill=black!22},
  data/.style={draw, align=center, inner sep=4pt, font=\small, fill=white},
  ghost/.style={data, dashed, fill=none, text=black!55, font=\small\itshape},
  disk/.style={data, double, double distance=1pt, fill=black!3},
  loopbox/.style={draw, dashed, rounded corners=4pt, inner sep=8pt},
  looplabel/.style={font=\scriptsize\itshape, anchor=south west, inner sep=1pt},
  stagegroup/.style={draw, semithick, rounded corners=5pt, inner sep=10pt},
  stagelabel/.style={font=\small\bfseries, anchor=south west, inner sep=2pt},
  arr/.style={-{Stealth[length=2.2mm]}, semithick},
}

\newcommand{\eps}{\varepsilon}
\newcommand{\ii}{\mathrm{i}}

\newcommand{\dd}{\mathrm{d}}

\renewcommand{\v}[1]{\bm{#1}}   

\journal{Computer Physics Communications}

\begin{document}

\begin{frontmatter}

\title{Fourier Transforms of Color Glass Condensate Multi-Wilson-Line Correlators via Filon Quadrature}
\author[a,b]{Haowu Duan}
\ead[]{haowu.duan@ccnu.edu.cn}
\author[a,b]{Si-Wei Dai}
\ead[]{swdai@mails.ccnu.edu.cn}
\author[a,b]{Cong Yi}
\ead[]{congyi@ccnu.edu.cn}
\author[a,b]{Wenbin Zhao}
\ead[]{WenbinZhao@ccnu.edu.cn}

\address[a]{Key Laboratory of Quark and Lepton Physics (MOE) and
Institute of Particle Physics, Central China Normal University, Wuhan
430079, China}
\address[b]{Artificial Intelligence and Computational Physics Research
Center, Central China Normal University, Wuhan 430079, China}

\begin{abstract}
Calculating cross sections in the Color Glass Condensate effective theory requires Fourier transforms of multi-Wilson-line correlators from transverse coordinate space to transverse momentum space. Under the common assumption of impact-parameter independence, each transform reduces to a set of Hankel transforms whose Bessel-function kernels oscillate rapidly at phenomenologically relevant momenta, making direct quadrature prohibitively expensive. We present a Filon-type quadrature, applicable to any integrand, that integrates these oscillatory factors in closed form on the stored coordinate grid, reducing each Hankel transform to a precomputed weight vector and the full nested transform chain to a sequence of matrix products. We develop and validate the method on the deep inelastic scattering dijet cross section beyond the correlation-limit approximation, where an exprel-based reformulation of the quadrupole Wilson-line correlator removes a numerical $0/0$ instability inherent to its standard parametrization. Porting the calculation to the Graphics Processing Unit (GPU), with custom CUDA kernels that fuse the momentum-space contraction directly into the correlator evaluation, brings the runtime for one dipole input down to about two minutes on a single NVIDIA A800, from several hours on a multi-core Central Processing Unit (CPU). We further generalize the algorithm to three sequential Hankel transforms and validate the resulting six-dimensional transform against an analytic Gaussian integrand family with closed-form results at every stage. This general, process-independent algorithm is directly applicable to next-to-leading-order proton-nucleus and electron-ion scattering cross-section calculations performed without the correlation-limit approximation. The code is publicly available at \url{https://github.com/CCNU-CGC-py/FFT_filon}.
\end{abstract}


\end{frontmatter}

\tableofcontents

\section{Introduction}
\label{sec:intro}
At high energy, the gluon density inside a hadron or nucleus grows rapidly and eventually saturates, giving rise to a new regime of strong-interaction dynamics governed by a large intrinsic momentum scale, the saturation scale $Q_s(x)$. The Color Glass Condensate (CGC) effective theory provides a systematic framework for this regime: the dense small-$x$ gluon fields are treated as classical color fields radiated by fast color sources, and scattering off these fields is encoded in Wilson lines that resum multiple interactions with the target to all orders in the strong coupling. Observables such as inclusive and semi-inclusive cross sections in deep-inelastic scattering (DIS) and in proton-nucleus collisions are then expressed through correlators of these Wilson lines~\cite{Kovchegov:2012mbw,McLerran:2002wj,Iancu:2003xm,Weigert:2005us,Gelis:2010nm,Albacete:2014fwa,Morreale:2021pnn}. Because the underlying dynamics is formulated in transverse coordinate space, where the small-$x$ evolution equations for the correlators take their simplest form, while physical cross sections are measured differentially in transverse momentum, cross sections in momentum space are obtained by Fourier transforming these correlators from transverse coordinate space to transverse momentum space. Numerically, this places nested two-dimensional Fourier transforms of multi-point correlators at the center of CGC phenomenology~\cite{Kharzeev:2004bw,Marquet:2007vb,Dominguez:2010xd,Dominguez:2011wm,Lappi:2012nh,Stasto:2018rci,Caucal:2025zkl}.

A common and often adequate approximation is to neglect the impact-parameter dependence of the correlators, which restores rotational invariance and reduces each two-dimensional transform to a set of Bessel (Hankel) transforms of angular modes. Even so, the oscillatory nature of the Bessel functions at the large gluon momenta of phenomenological interest remains a serious numerical obstacle. For dijet correlations, going beyond the back-to-back correlation-limit approximation requires solving for the full four-point correlator and Fourier transforming it from coordinate to momentum space, which arises already at leading order and in the large-$N_c$ limit  in DIS and in proton-nucleus collisions~\cite{Dominguez:2010xd,Dominguez:2011wm,Altinoluk:2011qy}. A few studies have tackled the numerical Fourier transform of the four-point Wilson-line correlator~\cite{Lappi:2012nh,Mantysaari:2019hkq}, but only at relatively low transverse momenta, where the Bessel oscillations are still mild enough for direct quadrature; at larger $p_T$ the cancellations between neighboring oscillations degrade any fixed-order scheme. The complementary strategy has been the \textit{Factorized Dipole Approximation}~\cite{Kovner:2017vro,Li:2021ntt,Kovner:2018azs,Duan:2022pma}, which simplifies the quadrupole into a product of dipoles and thereby bypasses the complicated Fourier transform. But this assumes that color fluctuations are independent and uncorrelated. It is valid only in the region $Q_s^2\,S_\perp\gg1$, with $S_\perp$ the effective transverse area of the proton/nucleus, and it discards the genuinely connected multi-gluon correlations that are important for gluon saturation effects. Extending full CGC calculations to the higher-$p_T$ region relevant for comparison with experimental data therefore requires an efficient and accurate method for evaluating these multi-point Wilson-line correlator transforms across the full kinematic range.

This work develops a quadrature designed for exactly this problem: a Filon-type rule, applied directly on the stored coordinate grid, in which the oscillatory factors are integrated in closed form. This collapses every Hankel transform into a precomputable weight vector, so that each transform, and the full nested chain built from them, reduces to a sequence of matrix products. The method is formulated for the general case of three nested transforms and is independent of the specific form of the integrand. We develop and validate the method on one concrete observable, the deep inelastic scattering dijet cross section at full angle and at fixed photon virtuality~\cite{Dominguez:2010xd,Dominguez:2011wm}, for two reasons. First, it offers an easier testing ground compare to the proton-nucleus case: only two of the three coordinate integrals carry a momentum transfer, so every intermediate result can be saved and inspected for validation. Second, the dijet computation beyond the correlation limit is of direct physical interest in its own right, and the tables produced here feed a CGC event generator described in a companion paper~\cite{haowu:2026}. The general three-momentum algorithm is nevertheless implemented in full and validated against an analytic integrand family with closed-form solutions.

The paper is organized as follows. Section~\ref{sec:dijet} sets up the dijet cross section, fixes the notation, and lays out the components of the calculation; Section~\ref{sec:cpu} assembles them into the CPU reference implementation; Section~\ref{sec:gpu} moves the expensive stage to the GPU, first keeping the inner table (Sec.~\ref{sec:tablecuda}) and then removing it altogether (Sec.~\ref{sec:fusedcuda}); Section~\ref{sec:threek} presents the extension of the algorithm to the general case of three sequential Hankel transforms, validated on an analytic family, with production use deferred to forthcoming proton-nucleus applications.

\section{The deep-inelastic scattering di-parton cross section}
\label{sec:dijet}

\subsection{Physical setup}
\label{sec:setup}
We first consider inclusive di-parton (dijet) production in deep-inelastic scattering, $\gamma^*(q_\gamma) + p/A \to q(k_1)\,\bar q(k_2) + X$, in the dipole picture of small-$x$ DIS. In light-cone perturbation theory the virtual photon of virtuality $Q^2$ splits into a quark-antiquark pair carrying light-cone momentum fractions $z$ and $1-z$ of the photon. The $q\bar q$ pair then scatters eikonally off the strong color field of the target, picking up Wilson-line phases, before fragmenting into the final state hadrons. Because at the leading order,  the splitting vertex appears once in the matrix element and once in its complex conjugate, the cross section involves the transverse dipole configuration $\v r$ in the matrix element and $\v r'$ in its complex conjugate, together with an impact-parameter variable $\v b$ that separates the two dipoles' centers. Eikonal scattering off the target correlates all four endpoints of these two dipoles, producing the four-Wilson-line (``quadrupole'') correlator $S^{(4)}$ that appears below. This process is also the standard tool for accessing the small-$x$ gluon Wigner distribution and gluon TMDs through dijet azimuthal correlations, and for constraining the dipole/quadrupole Wilson-line correlators that encode gluon saturation~\cite{Blaizot:2004wv,Fujii:2006ab,Dominguez:2011wm,Rezaeian:2012ye}.

Both photon transverse ($T$) and longitudinal ($L$) polarizations require the same type of integral, so let us denote the integrand as a function $G(\v{r}, \v{r}', \v{b}; \eps)$. In the case of DIS~\cite{Dominguez:2011wm},
\begin{equation}
\begin{split}
  I   =   \int\dd^2\v{r}\,\dd^2\v{r}'\,\dd^2\v{b}\,   e^{-\ii \v{k}_1\cdot(\v{r}-\v{r}')}e^{-\ii \v{q}\cdot \v{b}} \;   G_{T/L}(\v{r},\v{r}',\v{b};\eps),
\end{split}
\label{eq:physint}
\end{equation}
where
\begin{equation}
    \begin{split}
        G_{T}(\v{r},\v{r}',\v{b};\eps)&= \frac{\v{r}\cdot \v{r}'}{|\v{r}|\,|\v{r}'|} K_1(\eps |\v{r}|)K_1(\eps |\v{r}'|)\,S^{(4)}(\v{r},\v{r}',\v{b}), \\
        G_{L}(\v{r},\v{r}',\v{b};\eps)&= K_0(\eps |\v{r}|)K_0(\eps |\v{r}'|)\, S^{(4)}(\v{r},\v{r}',\v{b}),
    \end{split}
\label{eq:integrands}
\end{equation}
and the overall polarization prefactors $[(1-z)^2+z^2]\,\eps^2$ and $4Q^2\,z^2(1-z)^2$ multiply the transverse and longitudinal results once at the end. The modified Bessel functions $K_0,K_1$ and the scale $\eps^2=z(1-z)\,Q^2$ are the familiar output of the $\gamma^*\to q\bar q$ light-cone wavefunction. The scale $1/\eps$ sets the characteristic transverse size of the $q\bar q$ fluctuation. Throughout this work, momentum variables and scales, including $q$, $k_1$, $Q$, and $\eps$, are quoted in GeV, while transverse coordinate distances are quoted in $\mathrm{GeV}^{-1}$. The functions $K_0$ and $K_1$, with the angular factor $\v r\cdot\v r'/(|\v r||\v r'|)$ for $K_1$, arise from the spin structure of the longitudinally and transversely polarized photon vertices, respectively. The momentum $\v k_1$ is the transverse momentum of the tagged parton, while $\v q$, which is conjugate to $\v b$, is the net transverse momentum transferred to the target (the recoil of the $q\bar q$ pair as a whole). This differs from next-to-leading order forward dihadron production in $pA$ collisions, where two independent final-state momenta together with a fluctuating target impact parameter generally require Fourier transforms with respect to three independent transverse momenta~\cite{Chirilli:2012jd,Shi:2021hwx}. In DIS dijet production, by contrast, the cross section depends only on the relative dipole momentum $\v k_1$ and the pair recoil $\v q$. Under the assumption of impact-parameter independence, the target's overall impact parameter integrates out trivially, leaving only two Fourier conjugate pairs.

The four-point correlator is built from the dipole correlator $S(\v r)$, taken either from the Golec-Biernat--W\"usthoff (GBW) parametrization~\cite{Golec-Biernat:1998zce} or from a tabulated solution of small-$x$ evolution in which a neural network is used to extract an unparametrized initial dipole correlator, subsequently evolved with the collinearly-improved BK equation and fit globally to small-$x$ experimental data~\cite{Dai:2026nzp,Dai:2026rgl}. The resulting form is
\begin{equation}
\begin{split}
&S^{(4)}(\v{r},\v{r}',\v{b})
= -\frac{N}{D}\Big[S(\v r)S(\v r') - S(|\v{\Delta}+\v{b}|)\,S(\v b)\Big],\\
&N = F(\v{r}+\v{b},\v{b};\,0,\v{r}'),\qquad
D = F(\v{r}+\v{b},0;\,\v{b},\v{r}'),
\end{split}
\label{eq:color}
\end{equation}
where $\v{\Delta}=\v r-\v r'$, and $F(\v{x}_1,\v{x}_2;\v{x}_3,\v{x}_4)=\ln S(|\v{x}_1-\v{x}_3|)+\ln S(|\v{x}_2-\v{x}_4|)-\ln S(|\v{x}_1-\v{x}_4|)-\ln S(|\v{x}_2-\v{x}_3|)$. Physically, $S^{(4)}$ is the target average of four Wilson lines sitting at the endpoints of the two dipoles $\v r$ and $\v r'$, separated by $\v b$. It encodes the interference between the eikonal phase picked up in the matrix element and in its complex conjugate. Equation~\eqref{eq:color} is the Gaussian approximation and large-$N_c$ limit reduction of this quadrupole to products of dipole $S$-matrices, including the elastic and inelastic terms~\cite{Blaizot:2004wv,Fujii:2006ab,Dominguez:2011wm,Rezaeian:2012ye}. Only the two-point function $S(\v r)$, evolved by the BK equation, is needed as external input. The dimensionless combinations $N$ and $D$ implement this reduction exactly. Equation~\eqref{eq:color} takes the form $0/0$ when $D\to0$, or equivalently when $S(\v r)S(\v r')\to S(|\v{\Delta}+\v{b}|)S(\v b)$. We avoid this form using
\begin{equation}
\begin{split}
    S(\v r)S(\v r') - S(|\v{\Delta}+\v{b}|)S(b) = S(\v r)S(\v r')\big(1-e^{-D}\big)
\label{eq:exprel1}
\end{split}
\end{equation}
and as a result,
\begin{equation}
\begin{split}
    S^{(4)} = -N\,S(\v r)S(\v r')\,\texttt{exprel}(-D)
\end{split}
    \label{eq:exprel2}
\end{equation}
where $\texttt{exprel}(x)\equiv(e^x-1)/x$ and $\texttt{exprel}(0)=1$. For $|D|<10^{-7}$ we use $\texttt{exprel}(x)=1+x(\tfrac12+\tfrac{x}{6})$; otherwise we use the direct definition.

\subsection{Reduction of the integral for numerical evaluation}
\label{sec:numerics}
Here we focus on the integral involving only two Fourier transforms and exploit this structure to reduce it analytically before numerical evaluation.

For the given integrand $ G(\v{r},\v{r}',\v{b};\eps)$,
\begin{equation}
\begin{split}
  I   =   \int\dd^2\v{r}\,\dd^2\v{r}'\,\dd^2\v{b}\,   e^{-\ii \v{k}_1\cdot(\v{r}-\v{r}')}e^{-\ii \v{q}\cdot \v{b}}   G(\v{r},\v{r}',\v{b};\eps).
\end{split}
\end{equation}
with \(\v{\Delta}=\v{r}-\v{r}'\), i.e.\ \(\v{r}=\v{\Delta}+\v{r}'\), and define the relative angles
\begin{equation}
\begin{split}
  \alpha=\theta_\Delta-\theta_{r'},\qquad
  \beta=\theta_b-\theta_{r'}.
\end{split}
\end{equation}
Here $\theta_\Delta,\theta_{r'},\theta_b$ are the ordinary polar angles of $\v\Delta,\v r',\v b$, and $\alpha$, $\beta$ measure them relative to $\v r'$, which is the only vector shared by both phases below and whose overall orientation is therefore integrated out first. Writing $\widetilde G(\Delta,r',b,\alpha,\beta;\eps) =G(\v{\Delta}+\v{r}',\v{r}',\v{b};\eps)$, the integral becomes
\begin{equation}
\begin{split}
    I =& \int_0^\infty \Delta\,\dd\Delta
        \int_0^\infty r'\,\dd r'
        \int_0^\infty b\,\dd b
        \int_0^{2\pi}\dd\theta_{r'}
        \int_0^{2\pi}\dd\alpha
        \int_0^{2\pi}\dd\beta\; \\
        & \times
        e^{-\ii k_1 \Delta\cos(\alpha+\theta_{r'})}\,
        e^{-\ii q b\cos(\beta+\theta_{r'}-\theta_q)}\;
        \widetilde{G}(\Delta, r', b, \alpha, \beta;\eps),
        \end{split}
\end{equation}
There are only two Fourier phases. Expanding each gives
\begin{equation}
    \begin{split}
        e^{-\ii k_1 \Delta\cos(\alpha+\theta_{r'})}
        &= \sum_m (-\ii)^m\, J_m(k_1 \Delta)\, e^{\ii m\theta_{r'}}\, e^{\ii m\alpha}, \\
        e^{-\ii q b\cos(\beta+\theta_{r'}-\theta_q)}
        &= \sum_l (-\ii)^l\, J_l(q b)\, e^{\ii l\theta_{r'}}\, e^{\ii l\beta}\, e^{-\ii l\theta_q}.
    \end{split}
\end{equation}
The $\theta_{r'}$ integral enforces the following selection rule.
\begin{equation}
\begin{split}
    \int_0^{2\pi}\dd\theta_{r'}\; e^{\ii(m+l)\theta_{r'}} = 2\pi\,\delta_{m+l,\,0}
    \qquad\Rightarrow\qquad l = -m,
\end{split}
\end{equation}
while the angular Fourier modes are
\begin{equation}
\begin{split}
\widetilde G_{m,l}(\Delta,r',b;\eps) =   \int_0^{2\pi}\dd\alpha   \int_0^{2\pi}\dd\beta\,   e^{\ii m\alpha}e^{\ii l\beta}   \widetilde G(\Delta,r',b,\alpha,\beta;\eps).
\end{split}
\label{eq:modes}
\end{equation}
Using $\delta_{m,-l}$, and integrating out the radial variable \(r'\),
\begin{equation}
\begin{split}
  G_{m}(\Delta,b;\eps)   =   \int_0^\infty r'\,\dd r'\,   \widetilde G_{m,-m}(\Delta,r',b;\eps).
\end{split}
\label{eq:innertable}
\end{equation}
Here $G_{m}(\Delta,b;\eps)$ is the $m$-th azimuthal partial wave of the original integrand after the radial $r'$ integration, and should not be confused with $ G(\v{r},\v{r}',\v{b};\eps)$ itself. Introducing the double Bessel transform
\begin{equation}
\begin{split}
  {\cal I}_m   =   \int_0^\infty \Delta\,\dd\Delta   \int_0^\infty b\,\dd b\,   J_m(k_1\Delta)J_m(qb)\, G_{m}(\Delta,b;\eps),
\end{split}
\label{eq:Im}
\end{equation}
one obtains the two-Bessel radial form
\begin{equation}
\begin{split}
  I   =   2\pi   \sum_{m=-M}^{M}   (-1)^m e^{\ii m\theta_q}\,   {\cal I}_m.
\end{split}
\end{equation}
where the sum over the integer azimuthal harmonic $m$ is in principle infinite but is truncated at $|m|=M$, once ${\cal I}_m$ becomes negligible at the required numerical precision. In practice, $M$ is set by the smoothness of $G$ in the angles $\alpha,\beta$, i.e.\ by how quickly its angular Fourier coefficients fall off. Since ${\cal I}_m={\cal I}_{-m}$, caused by the real $S^{(4)}$ and  reflection symmetry of  $\alpha\to-\alpha,\beta\to-\beta$ of $\widetilde G$, this can be written as
\begin{equation}
\begin{split}
  I   =   2\pi\,{\cal I}_0   +   4\pi   \sum_{m=1}^{M}   (-1)^m\cos(m\theta_q)\,   {\cal I}_m.
\end{split}
\end{equation}
For the Hankel transform, we use Filon quadrature \cite{filon1930iii} that will be described next subsection. The scheduling of these contractions on Central Processing Unit (CPU) and Graphics Processing Unit (GPU), including memory management and custom CUDA kernel fusion, is described in the later sections.

\subsection{Filon--Bessel quadrature with closed-form local moments}
\label{sec:filon}
The double Bessel transform of Eq.~\eqref{eq:Im} is evaluated sequentially, with the $b$ integral performed first. Its smooth factor is $G_m(\Delta,b;\eps)$ at $\Delta$. Since the $\Delta$ dependence plays no role in the $b$ integral, we suppress it for brevity and write $G_m(b)$. The transforms $\int b\,J_m(qb)\,G_m(b)\,\dd b$ are oscillatory: at production-level momenta, $q\,b$ reaches $\mathcal{O}(10^3)$ across the $b$ grid, so any quadrature rule that directly resolves the oscillation requires a prohibitively fine mesh. Instead, we use a Filon-type rule~\cite{filon1930iii} on the \emph{stored} grid. On each interval $[b_i,b_{i+1}]$ the smooth function $G_m$ is fit by a degree-$d$ polynomial (default $d=4$) in the \emph{local} coordinate $u=b-b_i$,
\begin{equation}
G_m(b_i+u)\;\approx\;\sum_{r=0}^{d} c^{(i)}_r\,u^r,
\end{equation}
with $N_b$ nodes in total on the grid. The fit for interval $i$ uses the five points $L_i=\{b_{i-2},b_{i-1},b_i,b_{i+1},b_{i+2}\}$, while the resulting polynomial is integrated only over the interval $[b_i,b_{i+1}]$.

The oscillation is then integrated \emph{exactly} against each monomial via local moments, which admit a closed analytic form. Given the interval, with $b_i$ and $q$ held fixed, we expand the Bessel factor as a Taylor series in $u$ about $u=0$. The chain rule gives
\begin{equation}
J_m\big(q(b_i+u)\big)=\sum_{n\ge0}\frac{q^n J_m^{(n)}(qb_i)}{n!}\,u^n,
\end{equation}
where $J_m^{(n)}(qb_i)$ is the $n$-th derivative of $J_m$ evaluated at $q\,b_i$. On this interval, the Bessel function is thus replaced by a polynomial in $u$, which can be integrated against $u^r$ exactly. The elementary integral is
\begin{equation}
\int_0^{h_i} u^r\,\dd u=\frac{h_i^{r+1}}{r+1},
\label{eq:monomial}
\end{equation}
where $h_i=b_{i+1}-b_i$ is the width of the $i$th interval. Applying this term-by-term gives the moments
\begin{equation}
M_r^{(m)}(q;b_i)
= \int_0^{h_i} u^{r}\,J_m\big(q(b_i+u)\big)\,\dd u
= h_i^{r+1}\sum_{n\ge0}\frac{J_m^{(n)}(qb_i)}{n!}\,
\frac{(qh_i)^n}{r+n+1}.
\label{eq:moments}
\end{equation}
By Stirling's formula, $(q\,h_i)^n/n!\approx(e\,q\,h_i/n)^n$, so the terms in the series grow until $n\approx e\,q\,h_i$ and decay factorially beyond it. The required truncation order therefore scales with the interval width $h_i$. Since we evaluate all intervals with a single, common truncation order (to allow the local moments to be computed simultaneously through vectorization), this order must be set by the widest interval, $h_{\max}$, defined as the maximum value among $\{h_i\}$. We take $N=\lceil e\,q\,h_{\max}\rceil+30$, with the ceiling $\lceil \cdot \rceil$ rounds that estimate up to an integer, capped at $N\leq400$. Because $h_{\max}$ gives the slowest-converging case, this choice guarantees convergence on every interval, not just the widest one.

\subsection{Filon weight vectors}
\label{sec:weights}
The fit of Sec.~\ref{sec:filon} uses five sampled values on each interval. For interval $i$, let $j_0,\ldots,j_4\in L_i$ denote the global grid indices of these five fit points, and define the local offsets $u_k=b_{j_k}-b_i$ and sampled values $y_k=G_m(b_{j_k})$, for $k=0,\ldots,4$. The coefficients of the degree-4 polynomial satisfy
\begin{equation}
\sum_{r=0}^{4} c^{(i)}_r\,u_k^{\,r}=y_k,
\qquad k=0,\ldots,4.
\label{eq:conditions}
\end{equation}
These are five equations for the five unknown coefficients $c^{(i)}_0,\ldots,c^{(i)}_4$. Their coefficient matrix $V_{kr}=u_k^r$ is a Vandermonde matrix, with
\begin{equation}
\det V=\prod_{0\leq k<l\leq4}(u_l-u_k)\neq0,
\label{eq:vandermonde}
\end{equation}
and the five grid nodes are distinct. Equation~\eqref{eq:conditions} therefore has a unique solution.

Rather than solving Eq.~\eqref{eq:conditions} separately for every set of sampled values, we construct the solution using the Lagrange cardinal polynomials
\begin{equation}
\ell_k(u)=\prod_{\substack{l=0\\ l\neq k}}^{4}
\frac{u-u_l}{u_k-u_l},
\qquad k=0,\ldots,4.
\label{eq:cardinal}
\end{equation}
For example, the polynomial associated with the first fit point is
\begin{equation}
\ell_0(u)=
\frac{(u-u_1)(u-u_2)(u-u_3)(u-u_4)}
     {(u_0-u_1)(u_0-u_2)(u_0-u_3)(u_0-u_4)},
\label{eq:cardinalexample}
\end{equation}
and by construction, $\ell_k(u_l)=\delta_{kl}$. The unique interpolating polynomial is therefore
\begin{equation}
G_m(b_i+u)\approx\sum_{k=0}^{4} y_k\,\ell_k(u).
\label{eq:lagrange}
\end{equation}
Indeed, setting $u=u_l$ in Eq.~\eqref{eq:lagrange} kills every term except $k=l$, since $\ell_k(u_l)=\delta_{kl}$, leaving $y_l\ell_l(u_l)=y_l$. Uniqueness of this interpolant then follows from Eq.~\eqref{eq:vandermonde}.

To obtain the polynomial coefficients, we expand each cardinal polynomial in the monomial basis,
\begin{equation}
\ell_k(u)=\sum_{r=0}^{4}W^{(i)}_{rk}\,u^r.
\label{eq:fitmap}
\end{equation}
Inserting Eq.~\eqref{eq:fitmap} into Eq.~\eqref{eq:lagrange} gives $c^{(i)}_r=\sum_{k=0}^{4}W^{(i)}_{rk}y_k$. Relabeling each column by the corresponding global grid index gives the form used below,
\begin{equation}
c^{(i)}_r=\sum_{j\in L_i}W^{(i)}_{rj}\,G_m(b_j).
\label{eq:fitcoeffs}
\end{equation}
The matrix $W_{rj}^{(i)}$ depends only on the five node offsets, not on the sampled values, and is constructed once for each interval. In the implementation, the offsets are first divided by the width of the five-node window before solving the equivalent system, the $r$th coefficient is then divided by the $r$th power of that width. This rescaling keeps the system well conditioned as the grid is refined.

Combining Eq.~\eqref{eq:fitcoeffs} with the moments of Eq.~\eqref{eq:moments}, with $b=b_i+u$, the contribution of one interval is
\begin{equation}
\int_{b_i}^{b_{i+1}} b\,J_m(qb)\,G_m(b)\,\dd b
\;\approx\;
\sum_{r=0}^{d} c^{(i)}_r\,
\Big[\,b_i\,M^{(m)}_r(q;b_i)+M^{(m)}_{r+1}(q;b_i)\Big].
\label{eq:intervalsum}
\end{equation}
The two bracketed terms are the constant and varying parts of the Jacobian $b=b_i+u$, the constant $b_i$ multiplies the moment at the fit power, while the varying part $u$ raises the fit power by one and yields the moment one order up. With the fit coefficients still grouped, substituting Eq.~\eqref{eq:fitcoeffs} into Eq.~\eqref{eq:intervalsum} and summing over intervals gives 
\begin{equation}
\begin{split}
&\int_0^{b_{\max}} b\,J_m(qb)\,G_m(b)\,\dd b\; \\
\approx &\; \sum_{i}\,\sum_{r=0}^{d}\Big(\sum_{k\in L_i} W^{(i)}_{rk}\,G_m(b_k)\Big)\Big[\,b_i\,M^{(m)}_r(q;b_i)+M^{(m)}_{r+1}(q;b_i)\Big].
\label{eq:grouped}
\end{split}
\end{equation}
The parenthesized quantity is precisely $c^{(i)}_r$. Expanding the parenthesis and reordering the sums so that the samples appear first,
\begin{equation}
\begin{split}
&\int_0^{b_{\max}} b\,J_m(qb)\,G_m(b)\,\dd b \\
\;\approx&\; \sum_{i}\;\sum_{k\in L_i} G_m(b_k)\,\sum_{r=0}^{d} W^{(i)}_{rk}\,\Big[\,b_i\,M^{(m)}_r(q;b_i)+M^{(m)}_{r+1}(q;b_i)\Big]\\
\;=&\; \sum_{j=1}^{N_b} w_m^{j}(q)\,G_m(b_j).
\label{eq:weights}
\end{split}
\end{equation}
The second line merely regroups the first by grid node: each interval contributes five numbers, one to the weight-vector entry of each of its five fit points and since a node belongs to at most five fit sets, each entry $w_m^{j}(q)$ is the sum of at most five such contributions. Explicitly,
\begin{equation}
w_m^{j}(q)\;=\;\sum_{i:\,j\in L_i}\;\sum_{r=0}^{d}\,W^{(i)}_{rj}\,\Big[\,b_i\,M^{(m)}_r(q;b_i)+M^{(m)}_{r+1}(q;b_i)\Big],
\label{eq:wexplicit}
\end{equation}
where the outer sum runs over the (at most five) intervals whose fit set contains node $j$. Here no further index is needed, because the fit points in $L_i$ are labeled by their global grid indices, so the column $W^{(i)}_{rj}$ is defined whenever $j\in L_i$.

Every factor on the right of Eq.~\eqref{eq:weights} is now accounted for: $W^{(i)}$ depends only on the node positions, while the moments depend on the grid, the momentum $q$, and the order $m$. The integrand enters the transform only through its sampled values $G_m(b_j)$, and only linearly. The weight vector $w_m(q)$ is therefore a property of the (grid, momentum, order) triple, computed once and then reused for both polarizations and every dipole table on that grid. Stacking the rows over $m$ and over the target momenta turns each Hankel transform of the chain in Sec.~\ref{sec:numerics} into a single matrix product. The following sections describe how these products are scheduled on the CPU and GPU.

\section{The CPU reference implementation}
\label{sec:cpu}
The first production implementation runs on the CPU and is deliberately plain. Exact library functions evaluate the Bessel factors $K_0,K_1$, and a cubic spline in $r$ interpolates $\ln S$ through the dipole's tabulated nodes. Every later implementation is checked against this reference. Its structure is shared by all implementations and follows directly from the reduction of Sec.~\ref{sec:numerics}. The computation separates into two stages.
\begin{enumerate}
\item \textbf{Stage 1} evaluates the angular modes of Eq.~\eqref{eq:modes} and the $r'$ integral of Eq.~\eqref{eq:innertable}, producing the \emph{inner table} $G_m(\Delta,b;\eps)$ on the stored $(\Delta,b)$ grid. No momentum enters this stage.
\item \textbf{Stage 2} applies the two Hankel transforms of Eq.~\eqref{eq:Im} as Filon weight-vector contractions (Sec.~\ref{sec:weights}), turning the table into the angular modes of the cross section ${\cal I}_m(q,k_1)$. All momentum dependence lives here.
\end{enumerate}
The inner table connects the two stages: stage 1 writes it, stage 2 reads it, and nothing else crosses the boundary. Because the table is momentum-free, the expensive stage is paid once per dipole and $\eps$, while the cheap stage can be repeated on any momentum grid. Algorithm~\ref{alg:cpu} gives the pseudocode, and Fig.~\ref{fig:cpu} shows the workflow. The two stages are detailed below.

\begin{algorithm}[H]
\caption{The CPU reference implementation. Stage 1 builds the momentum-free inner table; stage 2 contracts it with Filon weight vectors; the inner table is the only object crossing the stage boundary.}
\label{alg:cpu}
\begin{algorithmic}[1]
\Statex \textbf{Stage 1 builds the inner table without momentum.}
\State precompute the $(\Delta,b)$ grids, angular sample points, the
$\ln S$ interpolation, Gauss--Legendre rule
$(r'_n,\omega_n)_{n=1}^{N_{r'}}$
\ForAll{$b$-chunks (one worker process per chunk)}
  \State evaluate the $r'$-free factors $\ln S(b)$,
  $\ln S(|\v{\Delta}+\v{b}|)$
  \For{$n=1,\dots,N_{r'}$}
    \State evaluate the guarded integrands of
    Eqs.~\eqref{eq:integrands} and \eqref{eq:exprel2} on the
    $(\Delta,b_{\text{chunk}},\alpha,\beta)$ block
    \State project onto the angular modes of Eq.~\eqref{eq:modes}; add with weight $\omega_n$
  \EndFor
\EndFor
\State multiply the polarization prefactors onto the finished arrays
\Statex\hrulefill
\Statex \textbf{Inner table} $G_m(\Delta,b;\eps)$ is held in memory,
archived on disk.
\Statex\hrulefill
\Statex \textbf{Stage 2 performs the momentum-dependent Hankel transforms.}
\State build the $\Delta$-grid weight rows $w^{(\Delta),l}_m(k_1)$ for
every $k_1$ \Comment{Sec.~\ref{sec:weights}}
\For{each $q$ of the momentum grid}
  \State build the $b$-grid weight rows $w^{(b),s}_m(q)$
  \State contract over $b$ and then over $\Delta$ using
  Eq.~\eqref{eq:twostep} for all $(k_1,m)$
\EndFor
\end{algorithmic}
\end{algorithm}

\subsection{From continuous to discrete}
\label{sec:cpu-build}
Every integral of Sec.~\ref{sec:dijet} is replaced by a finite sum. The angles are sampled on $N_\phi$ equally spaced points, $\alpha_i=2\pi(i{-}1)/N_\phi$, $i=1,\dots,N_\phi$, and likewise $\beta_j$. The $r'$ axis carries a Gauss--Legendre rule on $[0,R_{r'}]$ with nodes $r'_n$ and weights $\omega_n$, $n=1,\dots,N_{r'}$. The values $\Delta_l$, $l=1,\dots,N_\Delta$, and $b_s$, $s=1,\dots,N_b$ are the stored coordinate grids.

The angular Fourier transform, Eq.~\eqref{eq:modes}, becomes the equally spaced double sum. For a smooth $2\pi$-periodic integrand this sum is exact for every Fourier mode below the sampling limit, so the one requirement is $N_\phi>2M_{\max}$. With $l=-m$, the two phases combine into $e^{\ii m(\alpha-\beta)}$, and for the real integrands of Eq.~\eqref{eq:integrands} only its cosine survives. Applying the $r'$ rule to Eq.~\eqref{eq:innertable} then gives the discrete inner table on the stored grid points,
\begin{equation}
G_m(\Delta_l,b_s;\eps) =
\sum_{n=1}^{N_{r'}}\omega_n\,r'_n
\Big(\frac{2\pi}{N_\phi}\Big)^{2}
\sum_{i,j=1}^{N_\phi}
\cos\big(m(\alpha_i-\beta_j)\big)\,
\widetilde G(\Delta_l,r'_n,b_s,\alpha_i,\beta_j;\eps).
\label{eq:stage1discrete}
\end{equation}
The two Hankel transforms of Eq.~\eqref{eq:Im} are discretized by the Filon weight vectors of Sec.~\ref{sec:weights}, one axis at a time. The $b$ axis is transformed first, with rows built on the $b$ grid at momentum $q$, then the $\Delta$ axis, with rows built on the $\Delta$ grid at momentum $k_1$. Substituting the first contraction into the second,
\begin{equation}
{\cal I}_m(q,k_1) = \sum_{l=1}^{N_\Delta} w^{(\Delta),l}_m(k_1) \sum_{s=1}^{N_b} w^{(b),s}_m(q)\; G_m(\Delta_l,b_s;\eps).
\label{eq:twostep}
\end{equation}
The superscript in parentheses names the grid on which a weight row is built, and the accompanying index ($s$ on the $b$ grid, $l$ on the $\Delta$ grid) labels the grid node. The subscript $m$ is the Bessel order, common to both transforms by the selection rule of Sec.~\ref{sec:numerics}. The argument is the momentum at which the row was built. Every weight depends only on its (grid, momentum, order) triple, independent of the dipole and polarization, so the same rows serve both polarizations and every dipole table. Stacking the rows over the momentum grid then turns each of the two sums in Eq.~\eqref{eq:twostep} into a matrix product.

\subsection{The inner table}
\label{sec:cpu-table}
The product of stage 1 is a pair of real arrays (transverse and longitudinal) of shape $(M_{\max}{+}1)\times N_\Delta\times N_b$, and $\eps$ determines $R_\Delta$ and $N_\Delta$. The photon splitting wavefunction enters through $K_0(\eps|\v{r}|)$ and $K_1(\eps|\v{r}|)$, which vary on the scale $1/\eps$. As $\eps$ decreases, $K_1(\eps\, r)$ rises like $1/(\eps\, r)$ toward the origin, and the wavefunction reaches larger $|\v{r}|$ before dying off, so resolving it takes many points over a wider range. The $b$ direction is untouched by $\eps$ because the wavefunction depends only on $|\v{r}|$ and $|\v{r}'|$, so the $b$ profile of the table is set by the quadrupole correlator. The quadrupole falls off slowly at large $b$, as seen directly in Fig.~\ref{fig:innertable1}, so the $b$ grid is sized once for that long tail and reused unchanged for every $\eps$. Here we set $N_b=2048$ in every production table.

In production the range depends on $\eps$. We use $R_\Delta=15~\mathrm{GeV}^{-1}$ at $\eps\ge1~\mathrm{GeV}$, growing to $R_\Delta=107~\mathrm{GeV}^{-1}$ at the smallest production value $\eps=0.1~\mathrm{GeV}$. The $\Delta$ grid is uniform for $\eps\geq1~\mathrm{GeV}$; for $\eps<1~\mathrm{GeV}$ it uses the quadratic map $\Delta_j=R_\Delta[j/(N_\Delta-1)]^2$, which concentrates points near $\Delta=0$. The point count follows the range as $N_\Delta=\lceil200\sqrt{R_\Delta/(15~\mathrm{GeV}^{-1})}\,\rceil$,\footnote{On the quadratic grid, the first interval has width $R_\Delta/(N_\Delta-1)^2$. Scaling $N_\Delta$ as $\sqrt{R_\Delta}$ therefore keeps the fine spacing near $\Delta=0$ approximately unchanged while the outer boundary moves outward; the additional points mainly cover the newly included large-$\Delta$ tail.} giving $N_\Delta=200$ at the reference range and $N_\Delta=535$ at $\eps=0.1~\mathrm{GeV}$. The $r'$ integral widens and deepens alongside, $R_{r'}=60\to100~\mathrm{GeV}^{-1}$ and $N_{r'}=256\to384$, which raises the build cost but not the table size.

The resulting table size follows directly. With $M_{\max}=31$ and $N_b=2048$, the pair of arrays occupies disk storage about $0.2$\,GB at $\eps\ge1$ GeV and about $0.6$\,GB at $\eps=0.1$ GeV. The table is held in system memory and archived on disk as a plain-text table, one row per $(m,\Delta,b)$ entry with a small metadata file alongside. The text storage inflates the raw numbers roughly tenfold, so the largest production tables reach about $5$\,GB each on disk. The table is momentum-free, so angular modes of the cross section on a new momentum grid require only stage 2 to be rerun. A production set nevertheless spans many $(x,\eps)$ combinations at gigabytes each, and this storage burden is one of the motivations for the fused implementation of Sec.~\ref{sec:fusedcuda}, which produces the angular modes without ever materializing the table.

\begin{figure}[H]
\centering
\begin{tikzpicture}[node distance=3.5mm]
\node[data] (dip) {dipole $S(r)$ from GBW or tabulated small-$x$ evolution};
\node[cpu, below=9mm of dip] (spl) {$\ln S$ interpolation (cubic in $r$)};
\node[cpu, below=10mm of spl] (hoist) {$r'$-free factors
  $\ln S(b)$,\; $\ln S(|\v{\Delta}+\v{b}|)$};
\node[cpu, below=8mm of hoist] (eval) {evaluate integrands on
  $(\Delta,b_{\text{chunk}},\alpha,\beta)$\\
  exact $K_0,K_1$; exprel guard};
\node[cpu, below=of eval] (gemm) {angular projection onto the modes
  of Eq.~\eqref{eq:modes};\\ accumulate Gauss--Legendre weight};
\node[disk, below=13mm of gemm] (tab) {\textbf{inner table}
  $G_m(\Delta,b;\eps)$ is momentum-free,\\
  $0.2$--$0.6$\,GB in system memory (grows as $\eps$ decreases),
  archived on disk};
\node[cpu, below=13mm of tab] (bh) {$b$-Hankel using weight rows
  $w^{(b),s}_m(q)$, contract over $b$};
\node[cpu, below=of bh] (dh) {$\Delta$-Hankel using weight rows
  $w^{(\Delta),l}_m(k_1)$, contract over $\Delta$};
\node[data, below=9mm of dh] (am) {${\cal I}_m(q,k_1)$, both
  polarizations ($\sim$MB)};
\begin{scope}[on background layer]
\node[stagegroup, fit=(spl)(hoist)(eval)(gemm)] (s1) {};
\node[loopbox, draw=blue, fit=(eval)(gemm)] (rl) {};
\node[loopbox, draw=red, fit=(hoist)(rl)] (bl) {};
\node[stagegroup, fit=(bh)(dh)] (s2) {};
\end{scope}
\node[stagelabel] at (s1.north west) {Stage 1 builds $G_m$};
\node[looplabel] at (rl.north west) {for each $r'$ Gauss--Legendre node};
\node[looplabel] at (bl.north west) {process pool with one worker per $b$-chunk};
\node[stagelabel] at (s2.north west) {Stage 2 Hankel transforms};
\node[data, right=3mm of s2] (mom) {momentum\\ grids $(q,k_1)$};
\draw[arr] (dip) -- (spl);
\draw[arr] ([xshift=25mm]spl.south) -- ([xshift=25mm]bl.north);
\draw[arr] ([xshift=25mm]hoist.south) -- ([xshift=25mm]rl.north);
\draw[arr] (eval) -- (gemm);
\draw[arr] (s1.south -| tab) -- (tab);
\draw[arr] ([xshift=24mm]tab.south) -- ([xshift=24mm]s2.north -| tab);
\draw[arr] (bh) -- (dh);
\draw[arr] (s2.south -| am) -- (am);
\draw[arr] (mom) -- (s2);
\end{tikzpicture}
\caption{Workflow of the CPU reference implementation. Light rounded boxes are compute stages. Sharp rectangles are data, the double frame marks data saved to disk, and dashed rounded frames are loop scopes. Momentum never enters stage 1, and the momentum-free inner table is the only object that passes between them.}
\label{fig:cpu}
\end{figure}
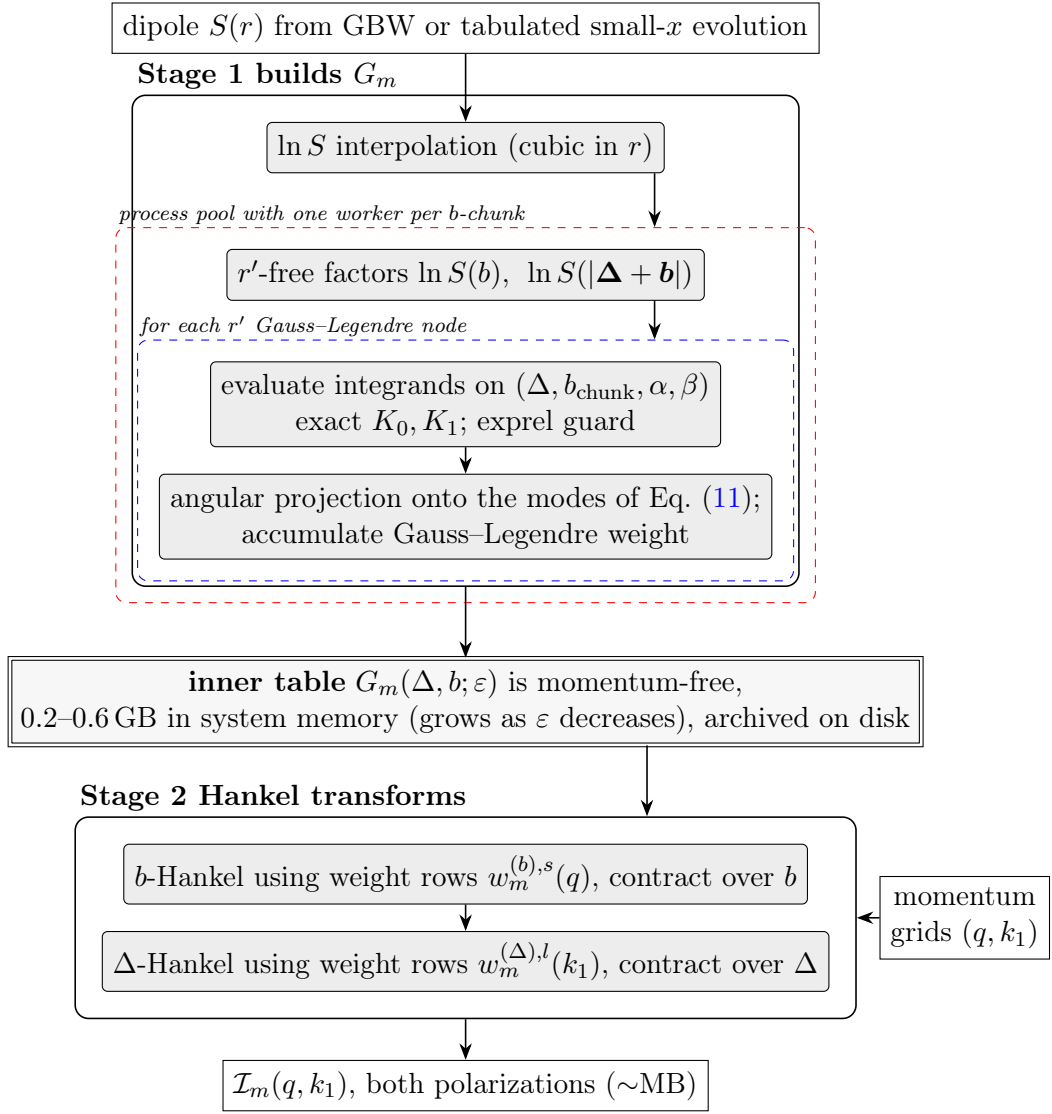

\subsection{Profile}
\label{sec:cpu-cost}
The wall time splits cleanly between the two stages. Table~\ref{tab:cpuprofile} reports the measured breakdown. All $\eps\ge1$ GeV share the same grid ($N_\Delta=200$, $N_{r'}=256$) and therefore the same cost, so a single measurement suffices. The $\eps=0.1$ GeV row is scaled from the measured per-point cost, which is constant per $r'$ node and per $b$-chunk across $\eps$. Only the time that needed to start the worker processes is not scaled.

\begin{table}[H]
\centering
\small
\caption{Wall-time breakdown of the CPU reference implementation on 16 cores ($N_b=2048$, $N_\phi=64$, $M_{\max}=31$, proton dipole at $x=0.01$, measured on 16 cores of a 20-core desktop).}
\label{tab:cpuprofile}
\begin{tabular}{lrrrrr}
\toprule
$\eps\ [\mathrm{GeV}]$ & $N_\Delta$ & $N_{r'}$ & stage 1 (16 cores) & stage 2 & total\\
\midrule
$\ge 1$ & 200 & 256 & $19\,943$\,s ($5.5$\,h) & 26\,s & $5.5$\,h\\
0.1 (scaled) & 535 & 384 & $80\,022$\,s ($22$\,h) & 69\,s & $22$\,h\\
\bottomrule
\end{tabular}
\end{table}

\subsection{Convergence and validation}
\label{sec:cpu-validation}
\begin{figure}[H]
\centering
\includegraphics[width=\textwidth]{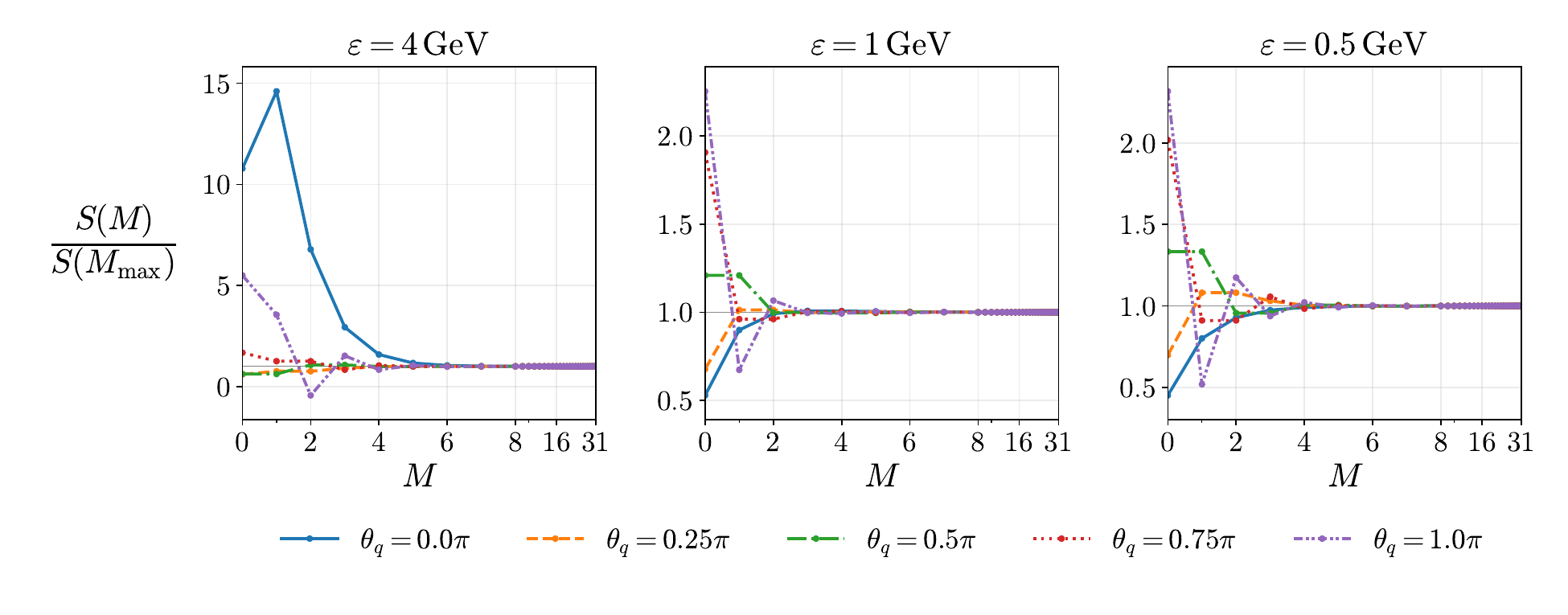}
\caption{Convergence of the angular mode sum $M$ for the proton dipole at $x=0.01$. The accumulated sum at $q=2~\mathrm{GeV}$, $k_1=5~\mathrm{GeV}$ is normalized by its converged value with $M_{\rm max} =31$, for five momentum angles $\theta_q$ and three photon scales $\eps$, quoted in GeV (transverse polarization). The horizontal axis is linear up to angular mode sum $M=8$ and logarithmic beyond.}
\label{fig:modeconv}
\end{figure}

Three convergence properties are checked. First, the angular mode sum of Sec.~\ref{sec:numerics} must saturate. The accumulated partial sum $\sum_{m=0}^{M}$ at fixed $(q,k_1)$ should flatten well before $M=M_{\max}$. Figure~\ref{fig:modeconv} demonstrates this for representative momenta.

Figure~\ref{fig:modeconv} establishes angular convergence at representative points, but the default CPU production resolution, $N_\phi=64$ and $M_{\max}=31$, does not resolve the entire small-$\eps$ region. At $\eps=0.5$ GeV, some production-grid points retain significant high-mode contributions and can even yield a negative truncated angular reconstruction. Resolving them requires increasing both the angular sampling and the retained mode count. Because the dominant stage-1 work scales as $N_\phi^2$, increasing $N_\phi$ from 64 to 256 multiplies the angular work by sixteen. Applied to the multi-hour and day-scale CPU costs in Table~\ref{tab:cpuprofile}, such a convergence study would require an unreasonable amount of CPU time. This computational limitation is one of the motivations for the GPU implementation.

We therefore defer the demanding small-$\eps$ test to Sec.~\ref{sec:gpu-angular-convergence}, after the GPU algorithm and its agreement with the CPU reference have been established. There we use the GPU to reach $N_\phi=256$ and $M_{\max}=127$ in a practical wall time and show explicitly that the same algorithm recovers a stable, positive result once sufficient angular resolution is supplied.

\begin{figure}[H]
\centering
\includegraphics[width=\textwidth]{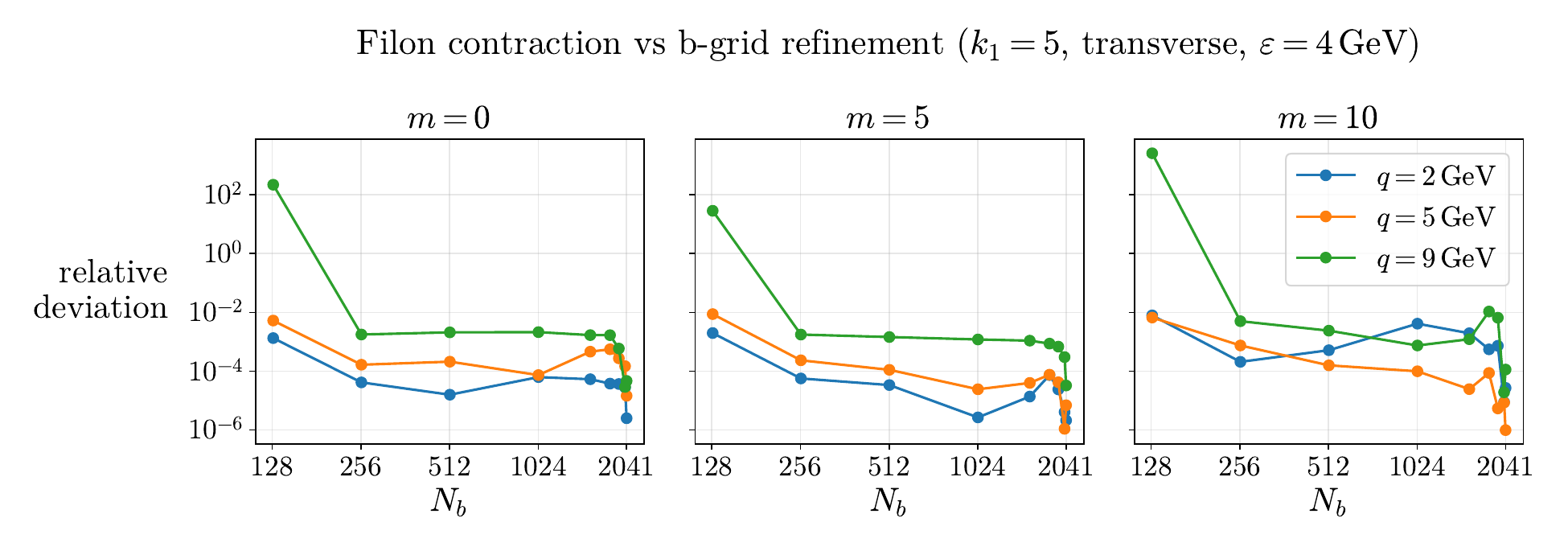}
\caption{Convergence of the Filon weight contraction with $b$-grid refinement for the proton table at $x=0.01$ and $\eps=4$ GeV. The curves show
the relative deviation of ${\cal I}_m(q,k_1{=}5~\rm GeV)$ from the full-grid ($N_b=2048$) result, for modes $m=0,5,10$ and momenta $q=2,5,9$ GeV. The coarse grids keep every 16th to every 2nd node; the finest grids remove every 4th to every 256th node, always keeping the endpoint so
the integration range is unchanged. At $N_b=2048$ the deviation is identically zero by construction.}
\label{fig:filonconv}
\end{figure}

Second, the Filon weight contraction must converge with the grid. At fixed momentum and mode, the transform computed on progressively finer grids (or with increasing fit degree) should settle to a stable value. Figure~\ref{fig:filonconv} shows the relative change as a function of grid refinement.

\begin{figure}[H]
\centering
\includegraphics[width=\textwidth]{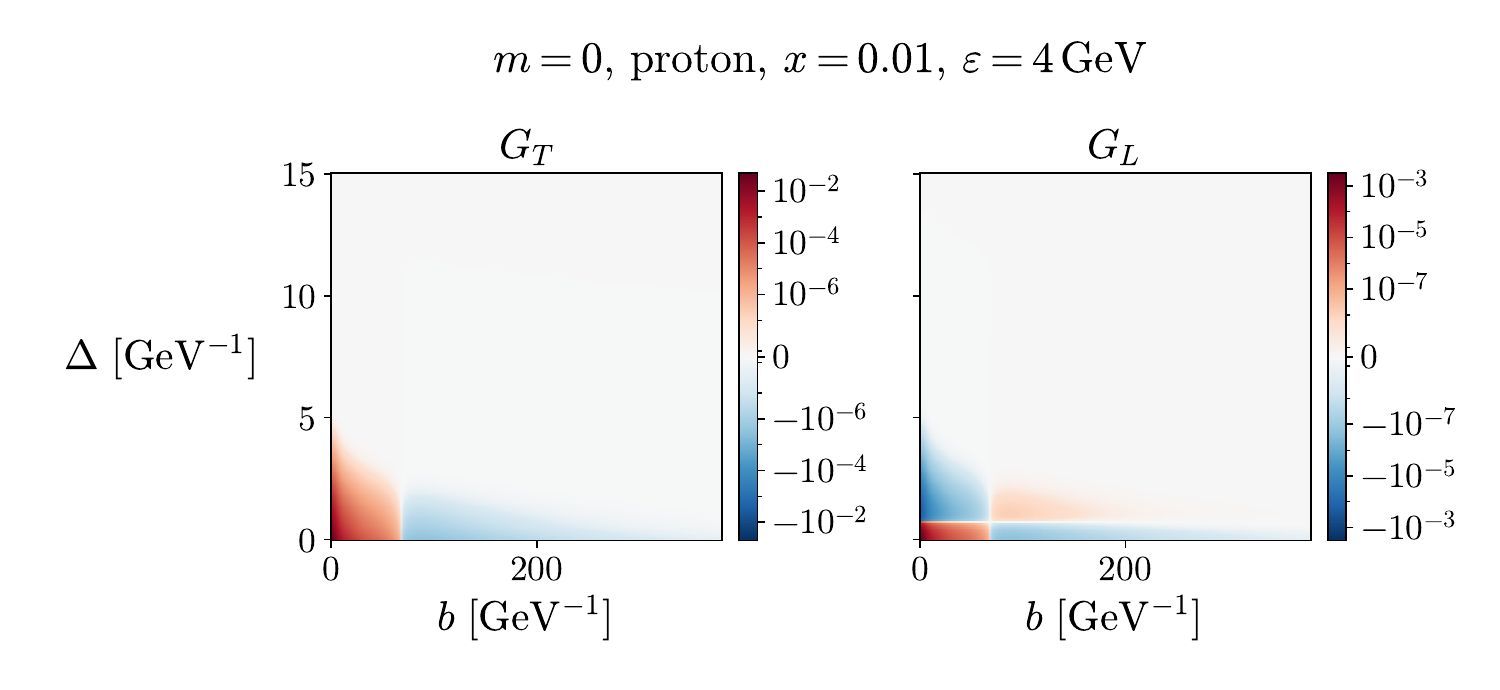}
\caption{Inner table $G_m(\Delta,b;\eps)$ for a proton dipole at $x=0.01$ and $\eps=4$ GeV, modes $m=0$, respectively. Each row contains the transverse (left) and longitudinal (right) polarizations. The color scale is symmetric logarithmic.}
\label{fig:innertable1}
\end{figure}

\begin{figure}[H]
\centering
\includegraphics[width=\textwidth]{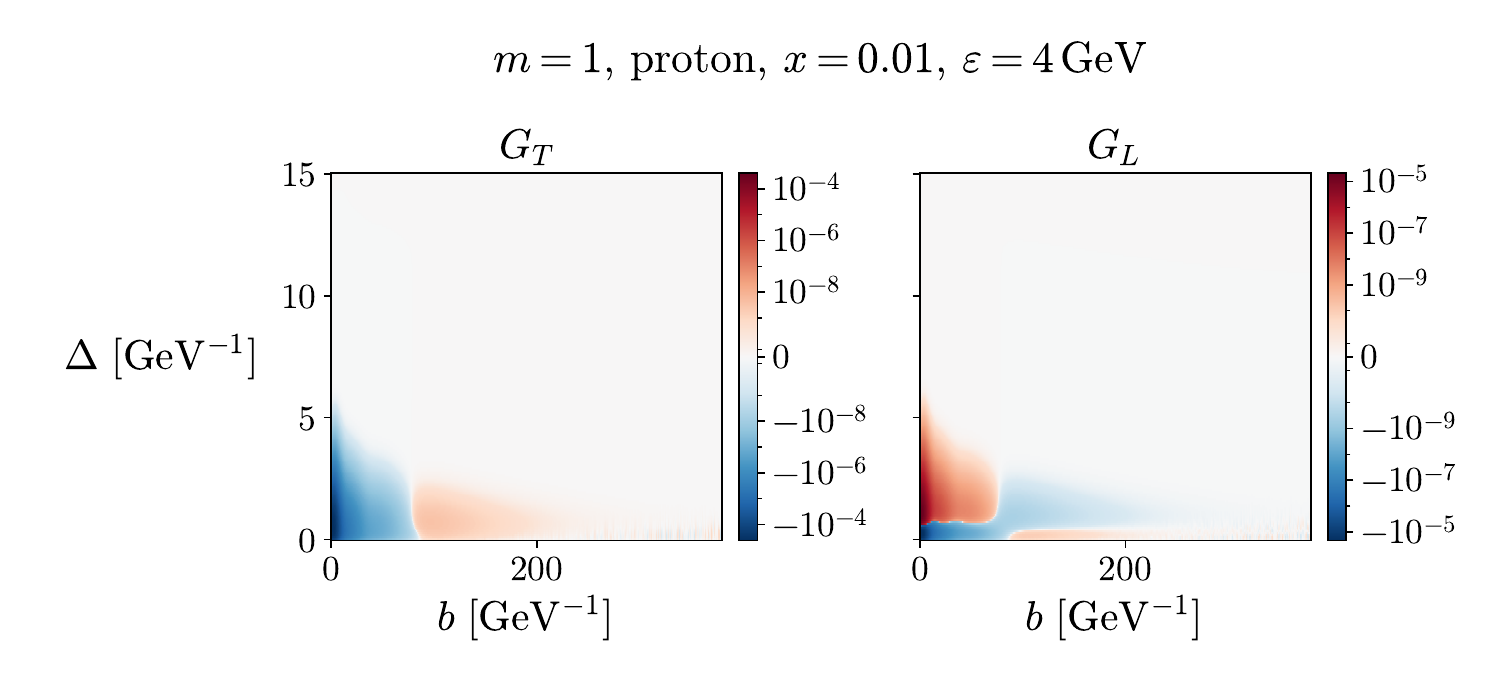}
\caption{Inner table $G_m(\Delta,b;\eps)$ for a proton dipole at $x=0.01$, $\eps=4$ GeV, and modes $m=1$, respectively. Each row contains the transverse (left) and longitudinal (right) polarizations. The color scale is symmetric logarithmic.}
\label{fig:innertable2}
\end{figure}

\begin{figure}[H]
\centering
\includegraphics[width=\textwidth]{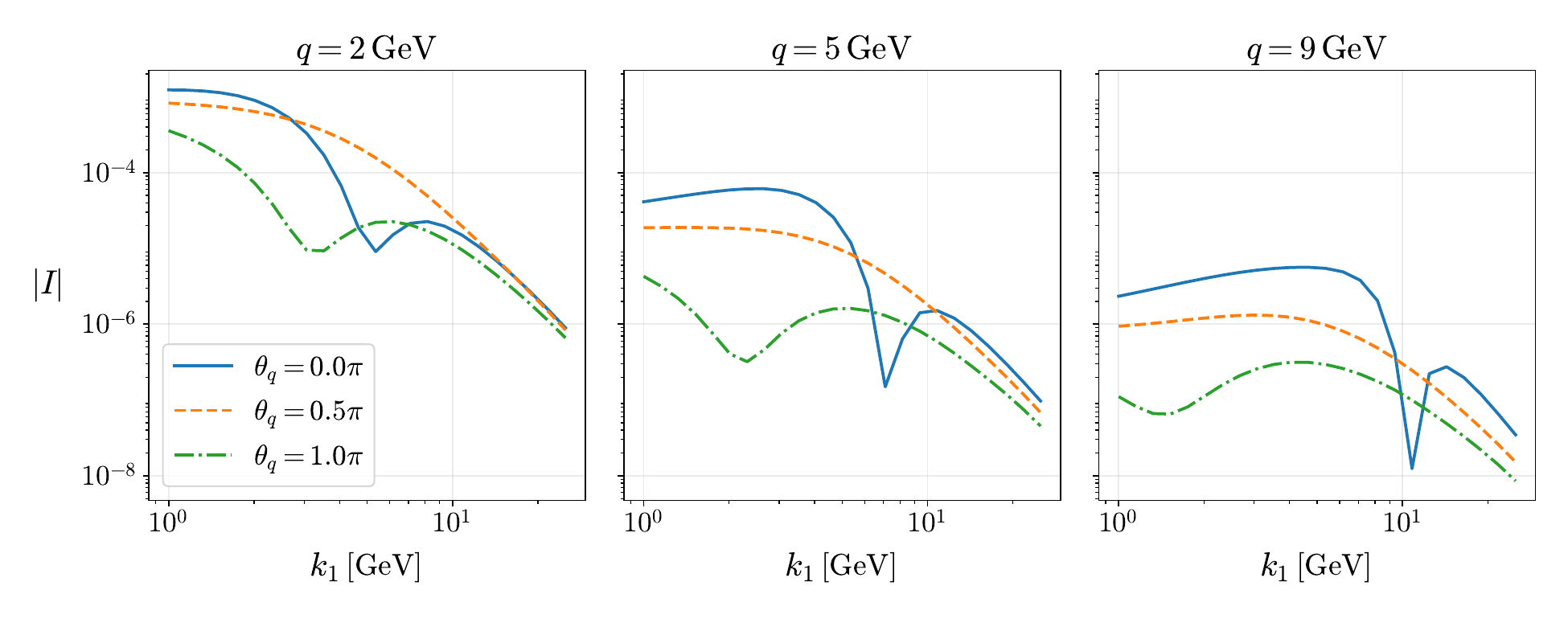}
\caption{Cross sections for a proton dipole at $x=0.01$, $\eps=4$ GeV in the transverse polarization. The folded mode sum $|I(q,k_1,\theta_q)|$ is plotted against $k_1$ at fixed $q=2,5,9$ GeV for three momentum angles on a log scale. The sharp dips are deep cancellation minima rather than zero crossings. The mode sum remains positive at every plotted point.}
\label{fig:trainingslice}
\end{figure}

Third, the inner table and the final cross sections are inspected visually. Fig~\ref{fig:innertable1}, and Fig~\ref{fig:innertable2} shows a heatmap of the inner table for a proton dipole, and Fig.~\ref{fig:trainingslice} shows slices of the resulting cross sections.


This reference implementation is the baseline against which all GPU implementations are checked. The following sections make the build itself fast by moving it to the GPU.

\section{The GPU implementations}
\label{sec:gpu}
Two GPU implementations are presented below.
Both leave the mathematics of the preceding sections untouched and use the same reduction, Filon weight vectors, and guarded integrand. The first (Sec.~\ref{sec:tablecuda}) keeps the two-stage structure of Sec.~\ref{sec:cpu} exactly and ports stage 1, which carries essentially all the cost, to CUDA for speed; the inner table remains a stored, reusable object. The second (Sec.~\ref{sec:fusedcuda}) goes one step further and removes the stage boundary itself: the Hankel contraction is a linear sum over table entries, so it can be applied to each piece of the table as that piece is computed, and the full table never needs to exist. The three-transform case developed later requires this capability.

\subsection{The CUDA table builder}
\label{sec:tablecuda}
The first GPU implementation is designed for speed. It computes the same inner table as Sec.~\ref{sec:cpu-table} and uses the same stage 2: that stage performs the weight contraction of Eq.~\eqref{eq:twostep} and runs on either processor. What changes is how stage 1 evaluates its $4\times10^{11}$ integrand samples. Two ideas provide the speedup.

First, the special functions are tabulated before the GPU calculation. Every transcendental ingredient is precomputed on the CPU as interpolation coefficients and evaluated on the GPU by polynomial interpolation. The first table stores $\ln S(r)$ on a uniform grid and is exact for the GBW dipole, whose $\ln S$ is a parabola in $r$. The regular combinations $xK_0(x)$ and $xK_1(x)$ (finite at $x=0$, where $K_0$ and $K_1$ themselves diverge) are stored in two additional tables. The tables are sized so that the lookup error is negligible compared to double precision.

Second, each factor of the integrand is computed only on the set of variables it actually depends on, instead of being recomputed at every point of the four-dimensional block. A naive kernel evaluates everything on $(\Delta,r',\alpha,\beta)$ for every $b$; here four custom kernels split the work as shown in Table~\ref{tab:kernelsplit}.
\begin{table}[H]
\centering
\small
\caption{Kernel decomposition of stage 1 on the GPU. The first three kernels evaluate factors reused in the final assembly of $S^{(4)}$.}
\label{tab:kernelsplit}
\setlength{\tabcolsep}{4pt}
\begin{tabular}{@{}
>{\raggedright\arraybackslash}p{0.20\textwidth}
>{\raggedright\arraybackslash}p{0.18\textwidth}
>{\raggedright\arraybackslash}p{0.12\textwidth}
>{\raggedright\arraybackslash}p{0.42\textwidth}@{}}
\toprule
kernel & depends on & frequency & computes\\
\midrule
wavefunction & $(\Delta,r',\alpha)$ & per $\Delta$-chunk &
$|\v{r}|=|\v{\Delta}+\v{r}'|$, $K_0(\eps|\v{r}|)$, $K_1(\eps|\v{r}|)$ via table lookup; $S(r)S(r')=e^{\ln S(r)+\ln S(r')}$\\[3pt]
$S(|\v\Delta+\v b|)S(b)$ & $(\Delta,\alpha,\beta)$ & per $b$ &
$\ln S(|\v{\Delta}+\v{b}|)$ and $\ln S(b)$; their sum and its exponential $S(|\v{\Delta}+\v{b}|)\,S(b)$\\[3pt]
$\ln S(|\v b-\v r'|)$ & $(\Delta,r',\beta)$ & per $b$ &
$\ln S(|\v{b}-\v{r}'|)$\\[3pt]
final $S^{(4)}$ assembly & $(\Delta,r',\alpha,\beta)$ & per $b$ &
$\ln S(|\v{r}+\v{b}|)$, $N$, $D$, and $S^{(4)}=-N\,S(r)S(r')\,\mathrm{exprel}(-D)$\\
\bottomrule
\end{tabular}
\end{table}
The only correlator evaluated is $S^{(4)}$, whose logarithmic combinations are
\begin{equation}
\begin{split}
N &= \ln S(|\v r+\v b|)+\ln S(|\v b-\v r'|)
     -\ln S(|\v\Delta+\v b|)-\ln S(b),\\
D &= \ln S(r)+\ln S(r')
     -\ln S(|\v\Delta+\v b|)-\ln S(b).
\end{split}
\label{eq:gpuND}
\end{equation}
The products $S(r)S(r')$ and $S(|\v\Delta+\v b|)S(b)$, together with $\ln S(|\v b-\v r'|)$, are evaluated on smaller domains and reused. The remaining term $\ln S(|\v r+\v b|)$ completes $N$ and depends on the full $(\Delta,r',\alpha,\beta)$ block at fixed $b$, so the ratio $N/D$ and the guarded result $S^{(4)}=-N\,S(r)S(r')\,\mathrm{exprel}(-D)$ are assembled only on that full block. This split avoids recomputing the reusable factors at every point. The angular projection then uses one dense matrix product per $b$, followed by the weighted $r'$ contraction.

This decomposition is specific to the DIS dijet integrand of Eq.~\eqref{eq:integrands}; a different observable would require its own kernel split, because the factorization exploits the particular dependence of the wavefunction and correlator on the integration variables.

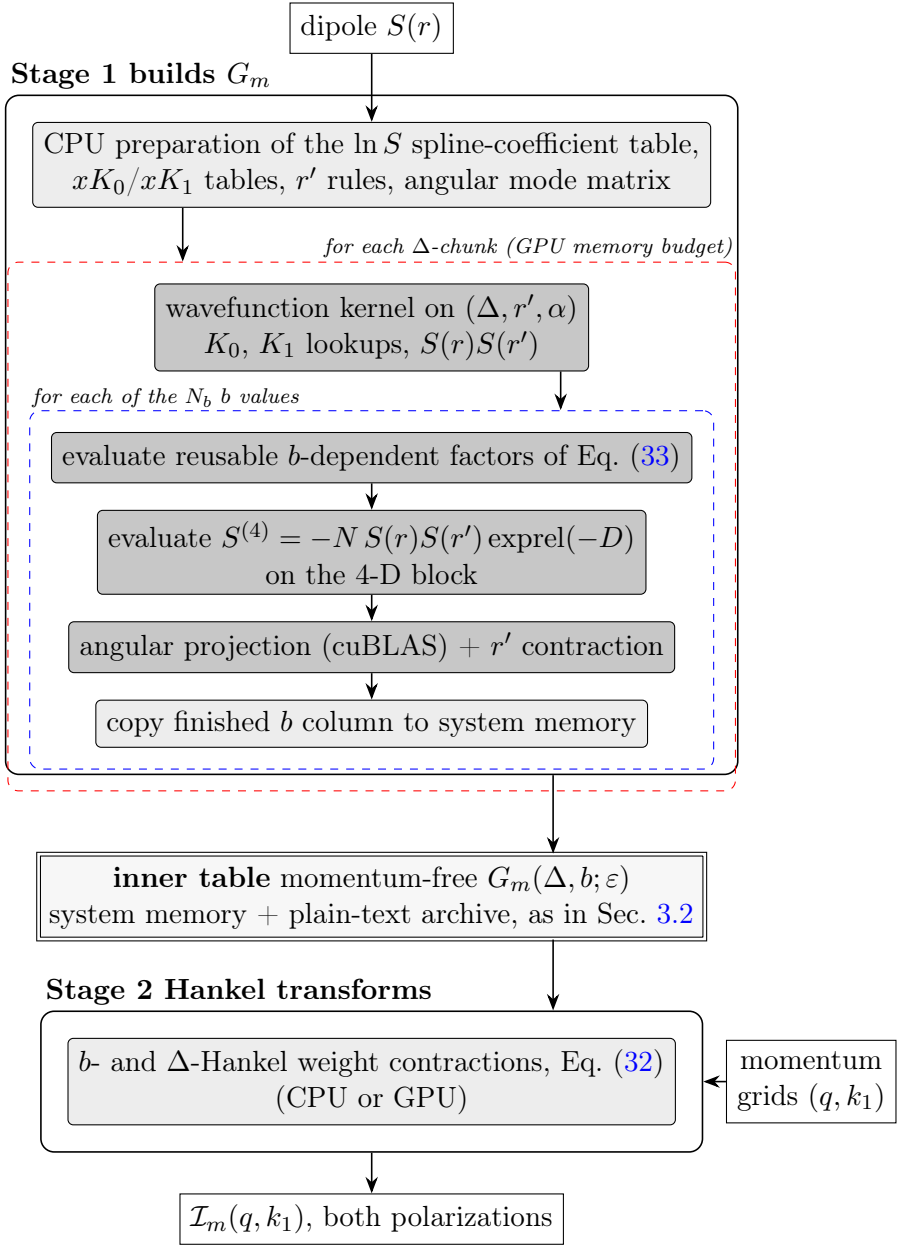
\begin{figure}[H]
\centering
\begin{tikzpicture}[node distance=3.5mm]
\node[data] (dip) {dipole $S(r)$};
\node[cpu, below=9mm of dip] (prep) {CPU preparation of the $\ln S$  spline-coefficient table,\\
  $xK_0$/$xK_1$ tables, $r'$ rules, angular   mode matrix};
\node[gpu, below=10mm of prep] (prer) {wavefunction kernel on   $(\Delta,r',\alpha)$\\
$K_0$, $K_1$ lookups, $S(r)S(r')$};
\node[gpu, below=8mm of prer] (preb) {evaluate reusable   $b$-dependent factors of Eq.~\eqref{eq:gpuND}};
\node[gpu, below=of preb] (main) {evaluate   $S^{(4)}=-N\,S(r)S(r')\,\mathrm{exprel}(-D)$\\
on the 4-D block};
\node[gpu, below=of main] (gemm) {angular projection (cuBLAS) + $r'$   contraction};
\node[cpu, below=of gemm] (copy) {copy finished $b$ column to system memory};
\node[disk, below=14mm of copy] (tab) {\textbf{inner table}
  momentum-free $G_m(\Delta,b;\eps)$\\ system memory + plain-text archive,
  as in Sec.~\ref{sec:cpu-table}};
\node[cpu, below=13mm of tab] (con) {$b$- and $\Delta$-Hankel weight
  contractions, Eq.~\eqref{eq:twostep}\\ (CPU or GPU)};
\node[data, below=9mm of con] (am) {${\cal I}_m(q,k_1)$, both
  polarizations};
\begin{scope}[on background layer]
\node[stagegroup, fit=(prep)(prer)(preb)(main)(gemm)(copy)] (s1) {};
\node[loopbox, draw=blue, fit=(preb)(main)(gemm)(copy)] (bloop) {};
\node[loopbox, draw=red, fit=(prer)(bloop)] (dloop) {};
\node[stagegroup, fit=(con)] (s2) {};
\end{scope}
\node[stagelabel] at (s1.north west) {Stage 1 builds $G_m$};
\node[looplabel] at (bloop.north west) {for each of the $N_b$ $b$ values};
\node[looplabel, anchor=south east] at (dloop.north east) {for each $\Delta$-chunk (GPU memory budget)};
\node[stagelabel] at (s2.north west) {Stage 2 Hankel transforms};
\node[data, right=3mm of s2] (mom) {momentum\\ grids $(q,k_1)$};
\draw[arr] (dip) -- (prep);
\draw[arr] ([xshift=-25mm]prep.south) -- ([xshift=-25mm]dloop.north);
\draw[arr] ([xshift=25mm]prer.south) -- ([xshift=25mm]bloop.north);
\draw[arr] (preb) -- (main);
\draw[arr] (main) -- (gemm);
\draw[arr] (gemm) -- (copy);
\draw[arr] ([xshift=24mm]s1.south -| tab) -- ([xshift=24mm]tab.north);
\draw[arr] ([xshift=24mm]tab.south) -- ([xshift=24mm]s2.north -| tab);
\draw[arr] (s2.south -| am) -- (am);
\draw[arr] (mom) -- (s2);
\end{tikzpicture}
\caption{Workflow of the CUDA table builder. Dark rounded boxes are CUDA stages; all other conventions are as in Fig.~\ref{fig:cpu}. The two-stage structure and the inner table between the stages are unchanged from the CPU implementation; only the production of the table moves to the GPU.}
\label{fig:tablecuda}
\end{figure}

Memory is scheduled around a GPU memory budget: the $\Delta$ axis is cut into chunks, and the two integrand buffers dominate, occupying about $2\times8\,N_{r'}N_\phi^2\approx17$\,MB per $\Delta$ row at production sizes. Each finished $b$ column of the projected table is copied to system memory when complete. The assembled table is stored in system memory and archived exactly as in Sec.~\ref{sec:cpu-table}: it remains momentum-free and reusable, but is produced faster. Algorithm~\ref{alg:tablecuda} gives the pseudocode, and Fig.~\ref{fig:tablecuda} shows the workflow; the stage separation of Fig.~\ref{fig:cpu} is unchanged.

\begin{algorithm}[H]
\caption{The CUDA table builder. Stage 2 is unchanged from Algorithm~\ref{alg:cpu}; only the production of the inner table moves to the GPU.}
\label{alg:tablecuda}
\begin{algorithmic}[1]
\Statex \textbf{Stage 1 builds the inner table with CUDA.}
\State prepare the grids, angular mode matrix, and $r'$ rules on the CPU; spline-coefficient tables for $\ln S$, $xK_0$, $xK_1$; upload to the GPU
\ForAll{$\Delta$-chunks (GPU memory budget)}
  \State wavefunction kernel on $(\Delta,r',\alpha)$ evaluates
  $K_0$, $K_1$ lookups, $S(r)S(r')$
  \For{each of the $N_b$ $b$ values}
    \State evaluate the reusable $b$-dependent factors of Eq.~\eqref{eq:gpuND}
    \State evaluate $S^{(4)}=-N\,S(r)S(r')\,\mathrm{exprel}(-D)$ on
    $(\Delta,r',\alpha,\beta)$ after the remaining $\ln S$ lookup
    \State angular projection (dense matrix product); $r'$ contraction
    \State copy the finished $b$ column to system memory
  \EndFor
\EndFor
\Statex\hrulefill
\Statex \textbf{Inner table} $G_m(\Delta,b;\eps)$ resides in system memory and is
archived on disk.
\Statex\hrulefill
\Statex \textbf{Stage 2 performs the Hankel transforms} as in
Algorithm~\ref{alg:cpu} (CPU or GPU).
\end{algorithmic}
\end{algorithm}

\subsection{Kernel profiling and wall-time breakdown}
\label{sec:tablecuda-profile}
The decomposition of Table~\ref{tab:kernelsplit} keeps the final $S^{(4)}$ assembly compute-bound rather than memory-bound; this is the only stage that touches the full four-dimensional block. Table~\ref{tab:ncuprofile} reports the hardware metrics measured by NVIDIA Nsight Compute for the five kernels at production grid sizes. The streaming multiprocessor (SM) and dynamic random-access memory (DRAM) entries are the achieved percentages of the corresponding limits \cite{NVIDIA:NsightCompute}. An SM value between 83 and 95 percent means that the processing resources are already operating near their sustained limit, while a DRAM value between 5 and 11 percent means that global-memory bandwidth remains far from saturation. In the Roofline interpretation, the large separation between these values identifies a compute-bound regime rather than a memory-bound one \cite{Williams:2009Roofline,He:2022Brrrr}: the available memory bandwidth is therefore sufficient, and the limiting factor is the hardware's compute capacity.

\begin{table}[H]
\centering
\small
\caption{Kernel profiling with NVIDIA Nsight Compute on an RTX\,3090 at production grid sizes ($\eps=4~\mathrm{GeV}$). SM and DRAM columns report the fraction of peak sustained throughput; all five kernels are compute-bound.}
\label{tab:ncuprofile}
\begin{tabular}{llrrr}
\toprule
kernel & bottleneck & SM\,[\%] & DRAM\,[\%] & occupancy\,[\%]\\
\midrule
wavefunction              & compute & 87 & 11 & 97\\
$S(|\v\Delta+\v b|)S(b)$ & compute & 83 & 8  & 93\\
$\ln S(|\v b-\v r'|)$    & compute & 86 & 8  & 94\\
final $S^{(4)}$ assembly   & compute & 87 & 10 & 96\\
angular projection (cuBLAS) & compute & 95 & 5  & 16\\
\bottomrule
\end{tabular}
\end{table}

Table~\ref{tab:gpuwalltable} reports the end-to-end wall time for one inner-table build on two GPUs. The NVIDIA RTX 3090 has a 1/64 FP64 rate, while the NVIDIA A800 has a full FP64 rate.\footnote{Here the FP64 rate is measured relative to peak FP32 throughput. A rate of $1/64$ means that peak FP64 throughput is one sixty-fourth of peak FP32 throughput. For the A800, ``full'' denotes its native scalar FP64 rate, for which peak FP64 throughput is one-half of peak FP32 throughput.} The breakdown into stage 1 and stage 2 is reported separately.

\begin{table}[H]
\centering
\small
\caption{Wall-time breakdown for one inner-table build on an RTX 3090 and an A800. Proton dipole at $x=0.01$. Only the stage-1 times are stable GPU measurements; stage 2 is dominated by the CPU construction of the Filon weight rows, so it reflects the CPU rather than the GPU and varies with CPU load. In production the weight rows are built once per $\eps$ and reused for every dipole, so the stage-2 cost is amortized.}
\label{tab:gpuwalltable}
\begin{tabular}{llrrr}
\toprule
GPU & $\eps\ [\mathrm{GeV}]$ & stage 1 & stage 2 & total\\
\midrule
RTX 3090  & 4.0 & 167\,s & 20\,s & 186\,s\\
RTX 3090  & 0.1 & 666\,s & 24\,s & 691\,s\\
A800      & 4.0 & 23\,s & 40\,s & 62\,s\\
A800      & 0.1 & 74\,s & 48\,s & 122\,s\\
\bottomrule
\end{tabular}
\end{table}

The percentages in Table~\ref{tab:ncuprofile} are normalized to the RTX~3090's own peak rates and do not represent the same absolute throughput on different GPUs. Stage~1 is dominated by double-precision arithmetic and is compute-bound, so the A800 accelerates this stage by providing much more FP64 throughput, which is the resource that limits the calculation. Table~\ref{tab:gpuwalltable} shows this directly: the stage-1 time decreases from 167 to 23 seconds at $\eps=4$ GeV and from 666 to 74 seconds at $\eps=0.1$ GeV, corresponding to speedups of about 7.3 and 9.0. Stage~2 does not follow this trend because its time is set mainly by the CPU construction of the Filon weights.

\begin{figure}[H]
\centering
\includegraphics[width=\textwidth]{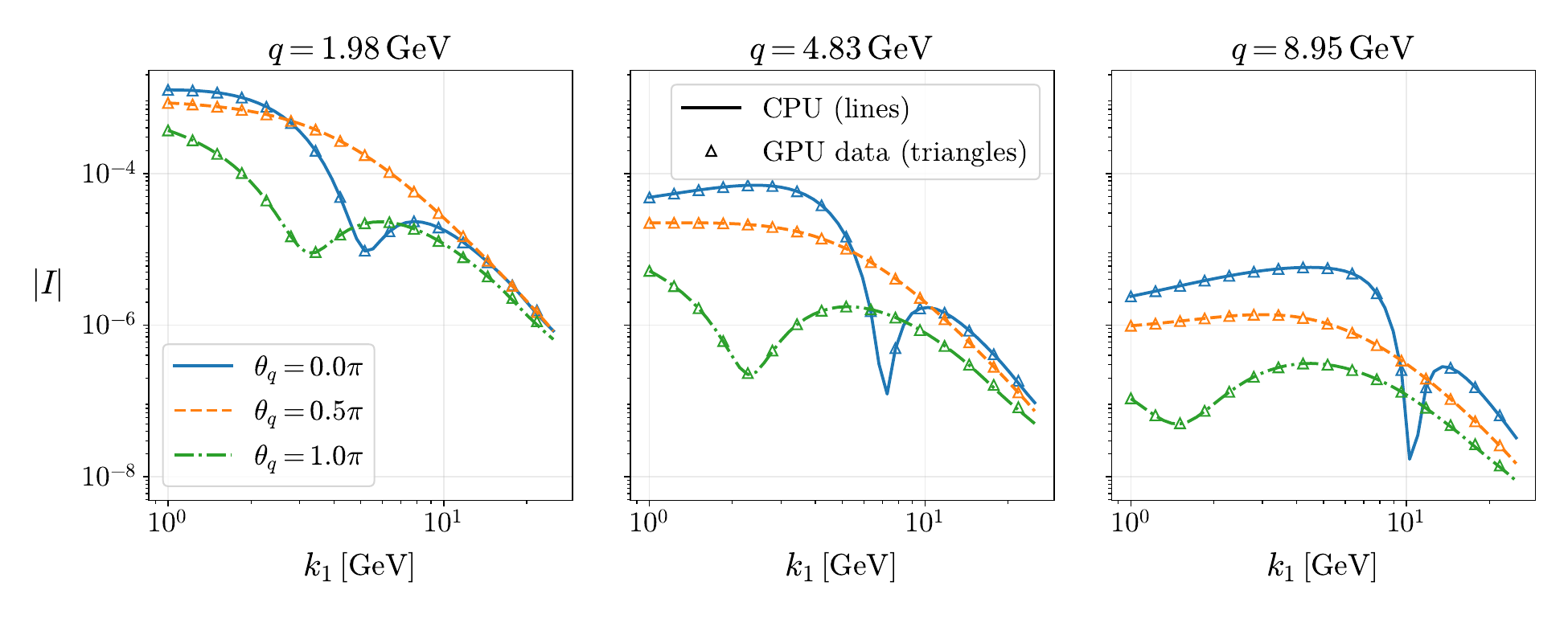}
\caption{Cross sections from the GPU table builder (open triangles, every third momentum point) overlaid on the CPU reference implementation (lines) for the proton dipole at $x=0.01$ and $\eps=4$ GeV in the transverse polarization. The quantity $|I(q,k_1,\theta_q)|$ is plotted against $k_1$ at three fixed values of $q$ for three momentum angles on a log scale.}
\label{fig:gpucheck}
\end{figure}

Figure~\ref{fig:gpucheck} closes the loop on correctness: the folded cross sections produced by the GPU table path lie on top of the CPU reference across momenta and angles, including through the deep cancellation minima of the log panels. The small residual discrepancies can be partially attributed to the improved treatment of the $r'$ integral in the GPU path relative to the CPU reference.

\begin{figure}[H]
\centering
\includegraphics[width=\textwidth]{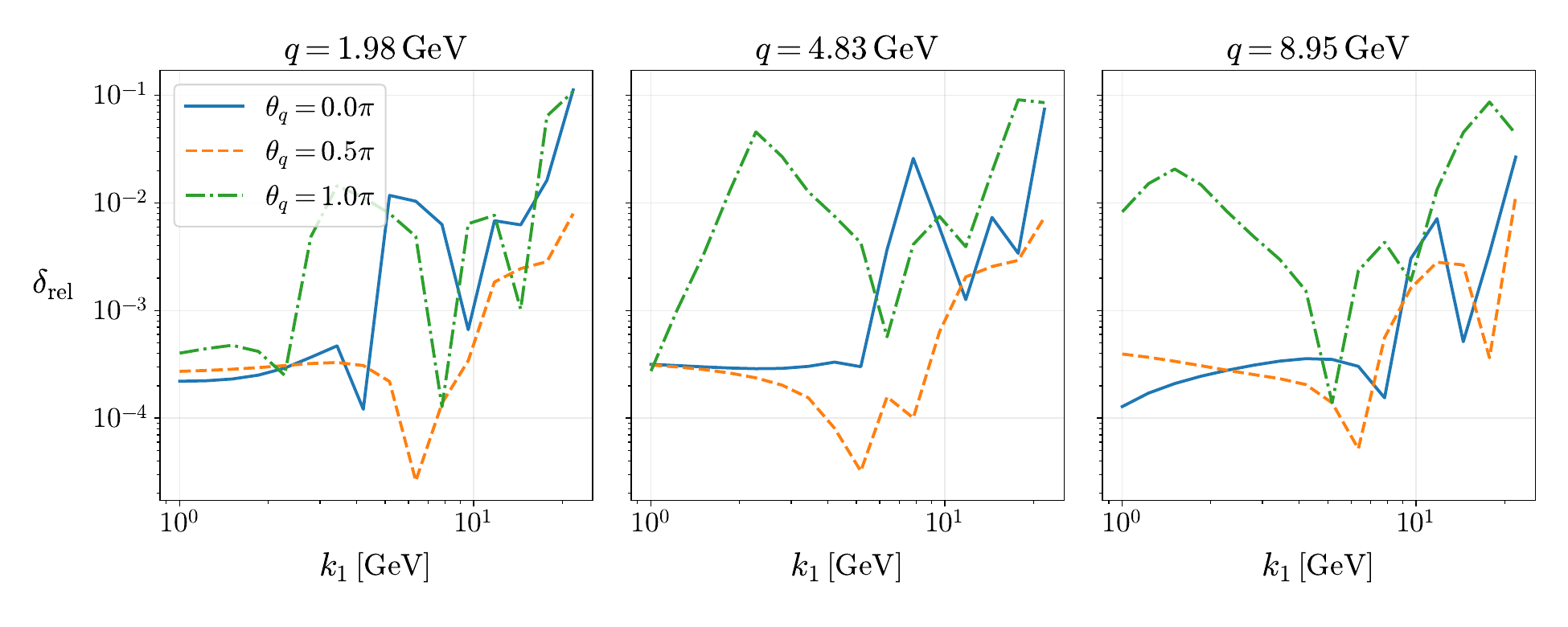}
\caption{Pointwise relative error corresponding to Fig.~\ref{fig:gpucheck}, defined as $\delta_{\mathrm{rel}}=\bigl||I_{\mathrm{GPU}}|-|I_{\mathrm{CPU}}|\bigr|/|I_{\mathrm{CPU}}|$. Only the GPU marker positions shown in Fig.~\ref{fig:gpucheck} (every third $k_1$ point) are used, and the curves connect these sampled values. The three panels use the same fixed $q$ values and the colors and line styles denote the same momentum angles $\theta_q$; both axes are logarithmic.}
\label{fig:gpucheck-relative-error}
\end{figure}

Figure~\ref{fig:gpucheck-relative-error} quantifies the same CPU--GPU agreement at the discrete GPU marker positions. The largest relative errors occur near the deep cancellation minima, where the CPU reference in the denominator becomes small.

\subsection{High-resolution angular convergence}
\label{sec:gpu-angular-convergence}
The comparison in Fig.~\ref{fig:gpucheck} establishes that the GPU implementation reproduces the CPU reference at the default production resolution. Its speed also makes it possible to increase the angular resolution far beyond what is practical in the CPU study. We now return to the small-$\eps$ limitation identified in Sec.~\ref{sec:cpu-validation} and use the GPU to test whether the apparent failure is a truncation effect or a failure of the algorithm itself.

\begin{figure}[H]
\centering
\includegraphics[width=\textwidth]{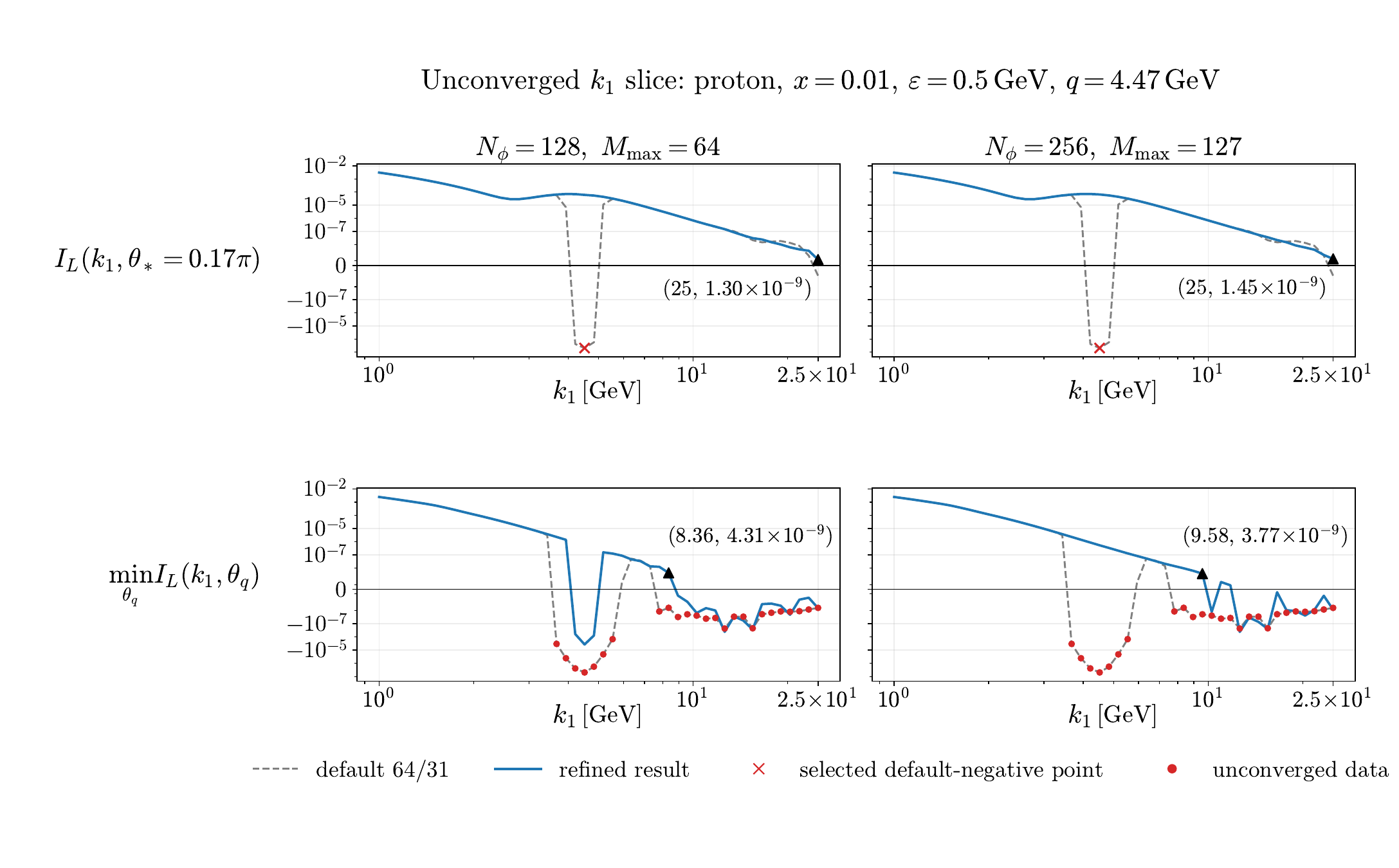}
\caption{Angular-resolution stress test for the proton dipole at $x=0.01$, $\eps=0.5$ GeV, and $q=4.47$ GeV.  The upper row shows the longitudinal mode sum as a function of $k_1$ at the fixed angle $\theta_*=0.17\pi$, where the selected default point is negative. The lower row shows, independently at each $k_1$, the smallest value over a dense 512-point scan of $0\leq\theta_q\leq\pi$; hence the minimizing angle need not be the same at different $k_1$.  The default $N_\phi=64$, $M_{\max}=31$ result is compared with $N_\phi=128$, $M_{\max}=64$ (left) and $N_\phi=256$, $M_{\max}=127$ (right).  Red points mark unconverged data in the default calculation.}
\label{fig:eps05-unconverged-slice}
\end{figure}

For the proton dipole at $x=0.01$ and $\eps=0.5$ GeV, replaying the angular scan at the default $N_\phi=64$, $M_{\max}=31$ resolution identifies an unconverged point at $q=4.47$ GeV, $k_1=4.51$ GeV, whose longitudinal result is most negative at $\theta_*=0.17\pi$. Figure~\ref{fig:eps05-unconverged-slice} shows the complete $k_1$ slice through this value of $q$. The GPU calculation permits us to repeat the same calculation at $N_\phi=128$, $M_{\max}=64$ and at the alias-safe resolution $N_\phi=256$, $M_{\max}=127$. As the angular resolution and retained mode count increase, the calculation progressively approaches the converged result: the $128/64$ result is already substantially improved, while the $256/127$ result recovers a positive angular minimum at the selected point.

\begin{figure}[H]
\centering
\includegraphics[width=\textwidth]{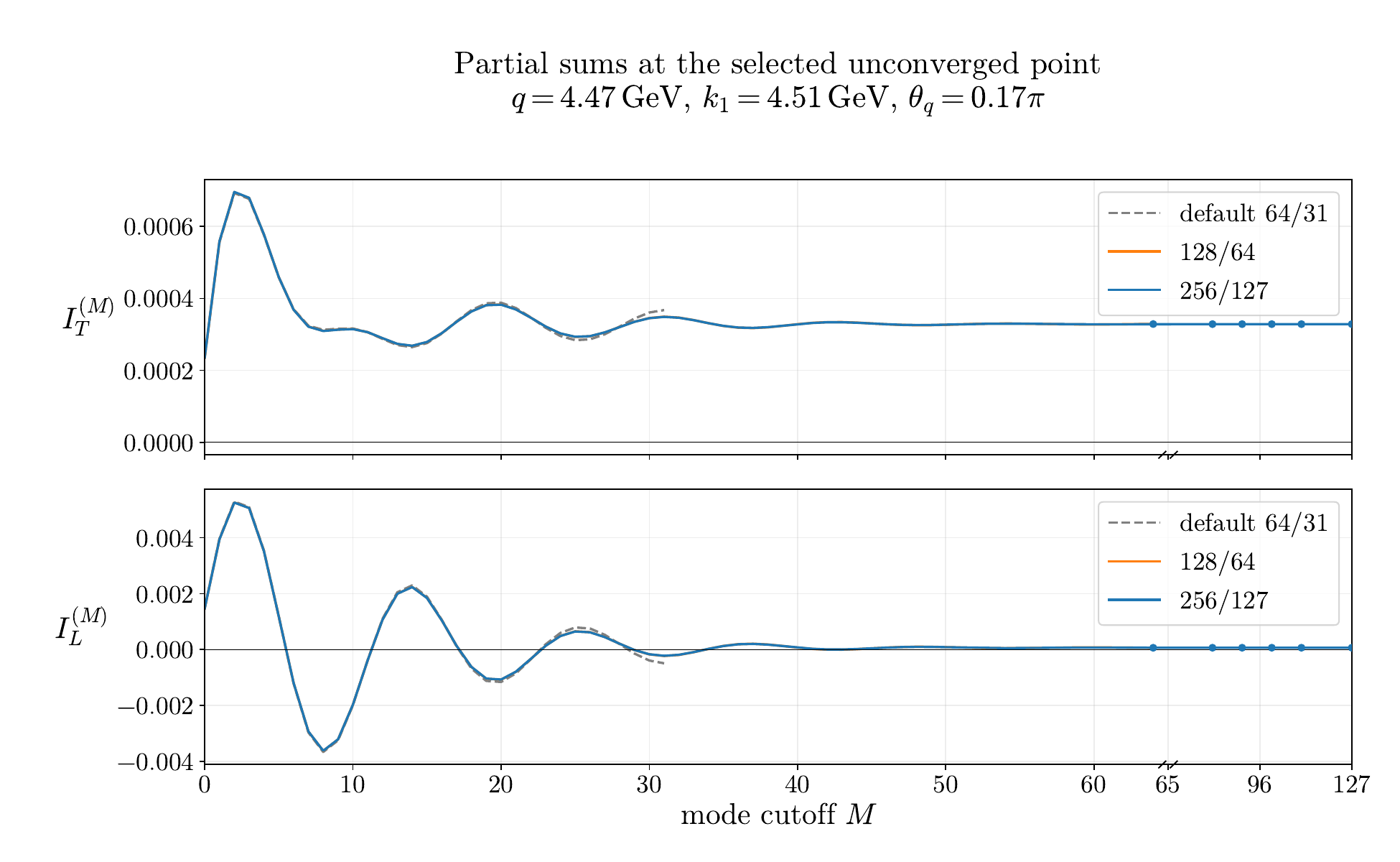}
\caption{Signed angular partial sums at the selected unconverged point $q=4.47$ GeV, $k_1=4.51$ GeV, and $\theta_q=0.17\pi$, for transverse (top) and longitudinal (bottom) polarization.  The default $N_\phi=64$, $M_{\max}=31$ curve is compared with the refined $N_\phi=128$, $M_{\max}=64$ and $N_\phi=256$, $M_{\max}=127$ calculations.  The circles on the high-resolution curve mark $M=64,80,90,100,110$, and 127.}
\label{fig:eps05-modeconv-signed}
\end{figure}

The recovery is seen more directly by following the partial mode sum at the selected point. Figure~\ref{fig:eps05-modeconv-signed} retains its sign and therefore displays both the oscillatory approach and the zero crossings. The default truncation terminates on the negative branch, whereas the finer angular FFT supplies the additional modes needed to reach the stable positive result. The $N_\phi=256$ calculation is performed only once, through $M_{\max}=127$. All intermediate cutoffs shown are partial sums of this same saved set of angular modes.

\begin{figure}[H]
\centering
\includegraphics[width=\textwidth]{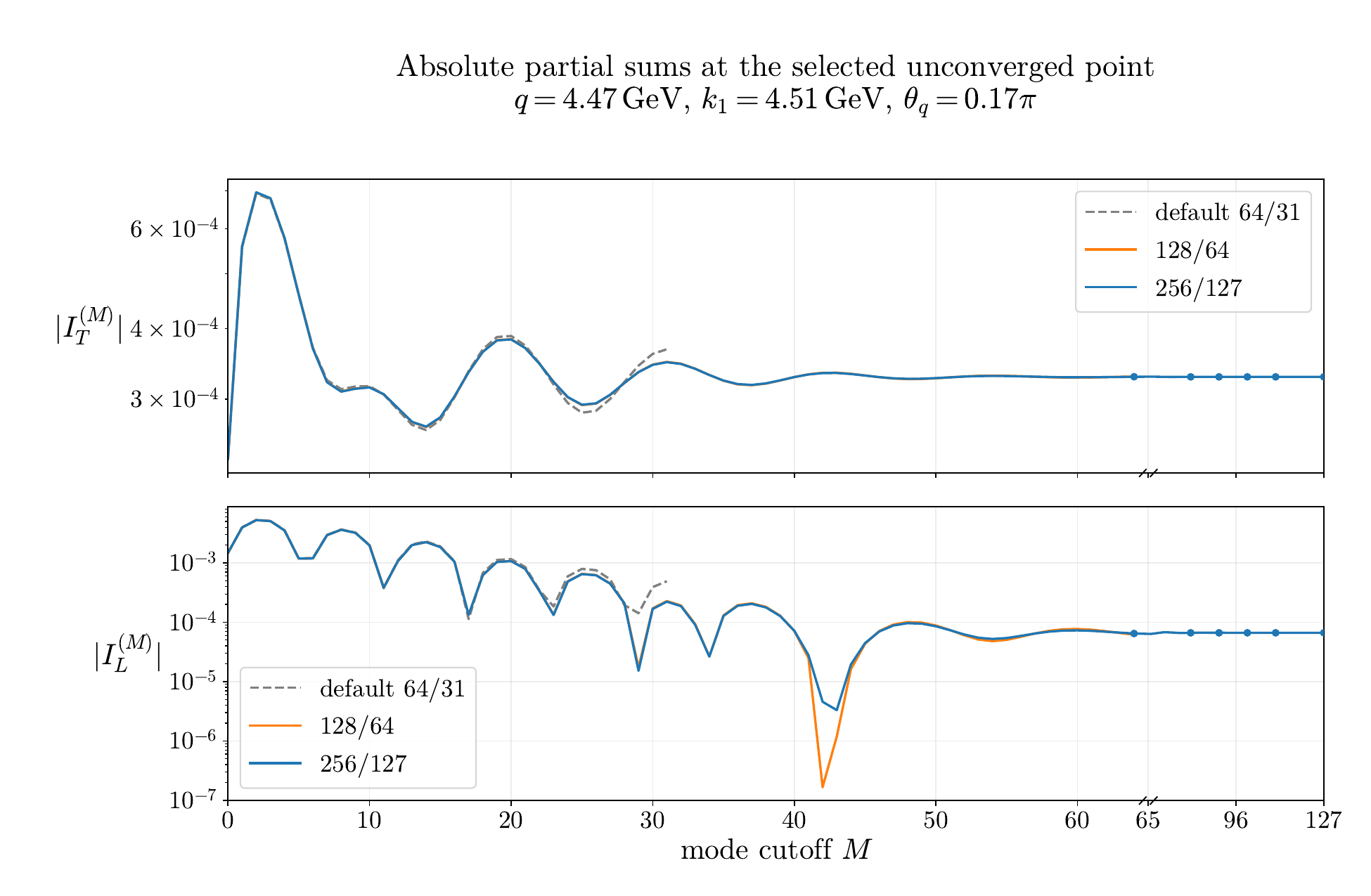}
\caption{Absolute values of the same partial sums as in Fig.~\ref{fig:eps05-modeconv-signed}, shown on a logarithmic vertical scale.  The decay of the longitudinal oscillations and the common high-$M$ plateau make the convergence of the refined angular calculation explicit.}
\label{fig:eps05-modeconv-log}
\end{figure}

Taking the absolute value exposes the tail over several orders of magnitude. Figure~\ref{fig:eps05-modeconv-log} shows that the oscillations damp and the result reaches a plateau. At the selected angle, the longitudinal partial sum at $M=110$ differs from the $M=127$ result by $8.0\times10^{-5}$ in relative terms. The GPU calculation therefore shows that the small-$\eps$ failure at the default resolution is purely an angular-truncation effect: the algorithm recovers the fully converged result once sufficient resolution is supplied.

\subsection{Computing the transforms without storing the inner table}
\label{sec:fusedcuda}

The dijet calculation can afford to store $G_m(\Delta,b;\eps)$ because this table has only two coordinate indices and one angular-mode index, and remains smaller than one gigabyte. The three-transform calculation discussed in Sec.~\ref{sec:intro} would add another coordinate index and another mode index; its table would be larger by orders of magnitude, and a separate table would be required for every combination of Bjorken $x$ and $\eps$. We therefore construct the transformed result without storing the inner table. The dijet calculation provides a setting in which this method can be tested before it is used for the three-transform case.

The starting point is an observation about Eq.~\eqref{eq:twostep}: both Hankel transforms are linear sums over table entries, so nothing requires the table to be complete before the sums begin. Split the contraction into its two steps,
\begin{equation}
\begin{split}
H_m(q,\Delta_l) &= \sum_{s=1}^{N_b} w^{(b),s}_m(q)\,
G_m(\Delta_l,b_s;\eps),\\
{\cal I}_m(q,k_1) &= \sum_{l=1}^{N_\Delta} w^{(\Delta),l}_m(k_1)\,
H_m(q,\Delta_l),
\end{split}
\label{eq:fused}
\end{equation}
and evaluate the first line as the values of $G_m$ are constructed: a small set of $G_m$ values is multiplied by the $b$ weights and added to $H$, and those values are then discarded. After all $b$ values have been included, the $\Delta$ weights are applied to $H$ to obtain ${\cal I}_m$. The full inner table is never formed or saved. This changes only the order of the additions, so the two implementations agree within rounding errors (Table~\ref{tab:fusedcheck}).

The grids, the $r'$ integration rules, the angular-mode matrix, the Bessel values, and the Filon weights are the same for every dipole at fixed $\eps$, and are therefore computed once; a new dipole requires only its $\ln S$ spline coefficients. The momenta $q$ and $k_1$ must be chosen when the Filon weights are computed, so changing these momenta requires repeating the calculation. This is the price of avoiding the stored table, which could otherwise be transformed again at new momenta.

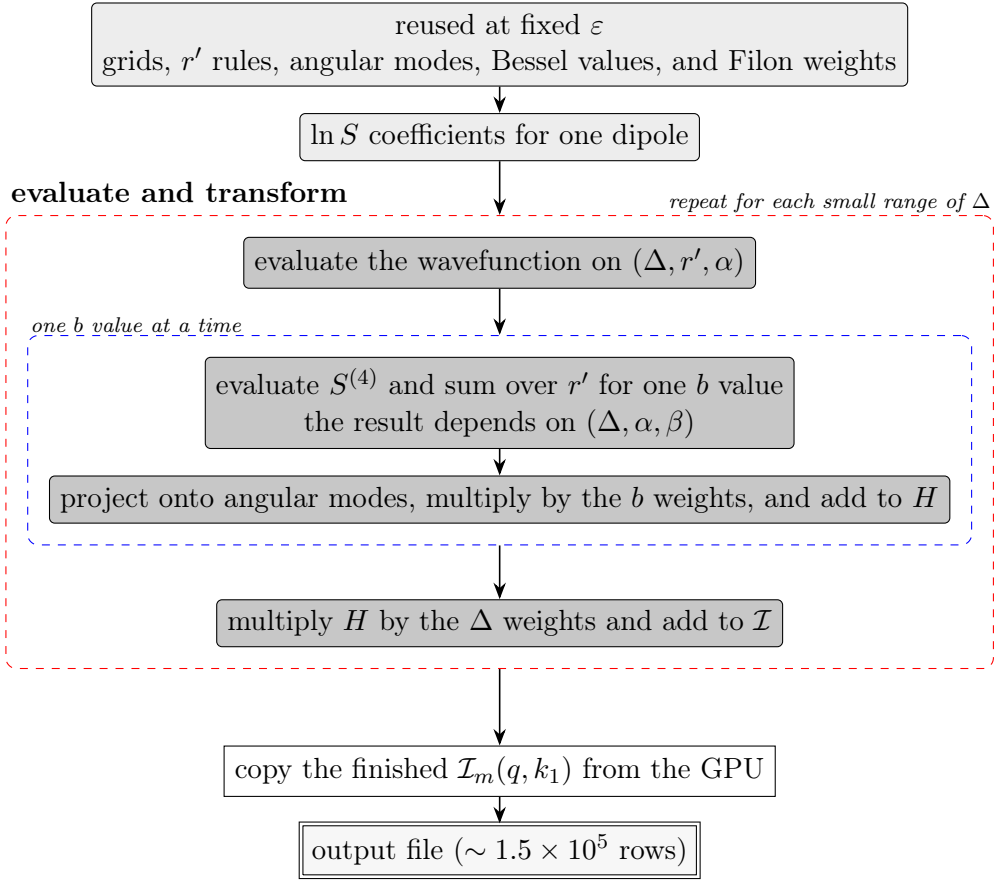
\begin{figure}[H]
\centering
\begin{tikzpicture}[node distance=3.5mm]
\node[cpu] (plan) {reused at fixed $\eps$\\ grids, $r'$ rules,
  angular modes, Bessel values, and Filon weights};
\node[cpu, below=of plan] (lns) {$\ln S$ coefficients for one dipole};
\node[gpu, below=10mm of lns] (prer) {evaluate the wavefunction on
  $(\Delta,r',\alpha)$};
\node[gpu, below=9mm of prer] (main)
  {evaluate $S^{(4)}$ and sum over $r'$ for one $b$ value\\ the result
  depends on $(\Delta,\alpha,\beta)$};
\node[gpu, below=of main] (hacc) {project onto angular modes, multiply
  by the $b$ weights, and add to $H$};
\node[gpu, below=10mm of hacc] (aacc) {multiply $H$ by the $\Delta$
  weights and add to ${\cal I}$};
\node[data, below=13mm of aacc] (am) {copy the finished
  ${\cal I}_m(q,k_1)$ from the GPU};
\node[disk, below=of am] (csv) {output file
  ($\sim1.5\times10^5$ rows)};
\begin{scope}[on background layer]
\node[loopbox, draw=blue, fit=(main)(hacc)] (bloop) {};
\node[loopbox, draw=red, fit=(prer)(bloop)(aacc)] (dloop) {};
\end{scope}
\node[stagelabel] at ([yshift=1mm]dloop.north west) {evaluate and transform};
\node[looplabel] at (bloop.north west) {one $b$ value at a time};
\node[looplabel, anchor=south east] at (dloop.north east) {repeat for each small range of $\Delta$};
\draw[arr] (plan) -- (lns);
\coordinate (fusedentry) at (dloop.north);
\draw[arr] (fusedentry |- lns.south) -- (fusedentry);
\draw[arr] (prer) -- (bloop.north -| prer);
\draw[arr] (main) -- (hacc);
\draw[arr] (bloop.south -| hacc) -- (aacc);
\draw[arr] (dloop.south -| am) -- (am);
\draw[arr] (am) -- (csv);
\end{tikzpicture}
\caption{Evaluation without storing the inner table. Each contribution is summed over $r'$ and transformed as it is evaluated. Only the finished result is copied from the GPU.}
\label{fig:fused}
\end{figure}

Figure~\ref{fig:fused} shows the calculation. It is repeated over the $\Delta$ grid, and the $b$ values are handled one at a time. For each $b$, $S^{(4)}$ is evaluated and summed over $r'$ in the same step, so the result depends only on $(\Delta,\alpha,\beta)$ and is projected directly onto the angular modes; multiplication by the $b$ weights then adds this contribution to $H$. After all $b$ values have been included, the $\Delta$ weights give ${\cal I}_m(q,k_1)$. The full $(\Delta,r',\alpha,\beta)$ integrand is never stored. At any point, the calculation stores only the current intermediate values, $H$ of order $10$\,MB, and the final result of a few MB; only the final result is copied from the GPU. The output contains about $1.5\times10^{5}$ rows instead of a table several gigabytes in size.

\subsection{Fused profiling and correctness}
\label{sec:fusedcuda-profile}

Table~\ref{tab:fusedwall} reports the total wall time of the fused implementation alongside the table-builder path of Sec.~\ref{sec:tablecuda} at the same $\eps$ and GPU. Because the fused path replays the same build kernels for the dipole correlator, any difference in total time reflects the cost of the incremental weight contraction versus the cost of writing and later reading the full table; the two are expected to be comparable.

\begin{table}[H]
\centering
\small
\caption{Wall time of the fused implementation vs the two-stage table builder on the same GPU and grid, proton dipole at $x=0.01$. The fused time is end-to-end and includes building the Filon weight rows. It is to be compared with the sum of the two table-builder stages. As in Table~\ref{tab:gpuwalltable}, stage 2 and the weight-construction part of the fused path are CPU-bound.}
\label{tab:fusedwall}
\begin{tabular}{llrr|r}
\toprule
GPU & $\eps\ [\mathrm{GeV}]$ & \multicolumn{2}{c|}{table builder} & fused\\
 & & stage 1 & stage 2 & \\
\midrule
RTX 3090  & 4.0 & 167\,s & 20\,s & 149\,s\\
RTX 3090  & 0.1 & 666\,s & 24\,s & 279\,s\\
A800      & 4.0 & 23\,s & 40\,s & 49\,s\\
A800      & 0.1 & 74\,s & 48\,s & 84\,s\\
\bottomrule
\end{tabular}
\end{table}

Because the fused path folds the $r'$ sum into the build kernel and processes one $b$ node at a time, the order in which the same finite sums are accumulated differs from that of the table builder. Table~\ref{tab:fusedcheck} reports the maximum absolute and relative differences between the fused and table-builder angular modes of the cross section ${\cal I}_m(q,k_1)$ across all modes, momenta, and both polarizations.

\begin{table}[H]
\centering
\small
\caption{Agreement between the two GPU paths at production grid sizes. Both paths compute the angular modes $\mathcal{I}_m(q,k_1)$ of the cross section for the proton dipole at $x=0.01$ on identical $48\times48$ momentum grids with geometrically spaced $q,k_1\in[1,25]~\mathrm{GeV}$. The table builder stores the full inner table and contracts it with the Filon weights afterwards. The fused path contracts on the fly with the $r'$ sum folded into the build kernel. Quoted are the largest absolute and relative entry-wise differences over all $48\times48\times32$ angular modes and both polarizations.}
\label{tab:fusedcheck}
\begin{tabular}{llll}
\toprule
$\eps\ [\mathrm{GeV}]$ & GPU & max abs diff & max rel diff\\
\midrule
4.0 & RTX 3090 & $2\times10^{-18}$ & $6\times10^{-5}$\\
0.1 & RTX 3090 & $4\times10^{-13}$ & $7\times10^{-4}$\\
4.0 & A800 & $3\times10^{-18}$ & $7\times10^{-5}$\\
0.1 & A800 & $1\times10^{-12}$ & $7\times10^{-4}$\\
\bottomrule
\end{tabular}
\end{table}

\section{The six-dimensional Fourier transform}
\label{sec:threek}
The dijet computation in the previous sections uses only two of the three sequential Hankel transforms, appropriate for the leading-order deep-inelastic-scattering process. This section presents the algorithm for the general case and validates it on an analytic family. Its production deployment is deferred to forthcoming next-to-leading-order proton--nucleus applications, where the third momentum is physical.

The general problem attaches an independent momentum to each of the three working variables,
\begin{equation}
\mathcal F(\v{k},\v{p},\v{q}) = \int \dd^2\v{\Delta}\,\dd^2\v{r}'\,\dd^2\v{b}\;
e^{+\ii(\v{k}\cdot\v{\Delta}+\v{p}\cdot\v{r}'+\v{q}\cdot\v{b})}\, \mathcal G(\Delta,r',b,\alpha,\beta),
\label{eq:sixdim}
\end{equation}
which is a genuinely six-dimensional Fourier transform, evaluated as three sequential Hankel transforms. The fused implementation of Sec.~\ref{sec:fusedcuda} was designed for precisely this case: the momentum-free intermediate now carries three coordinate axes and two mode indices, far too large to store per Bjorken-$x$ and $\eps$, so only the fused path is viable. The algorithm does not require a specific form of $\mathcal G$. We test it below with an integrable function whose transform is known exactly.

\subsection{Angular reduction with three phases}
\label{sec:threek-modes}
The angular reduction proceeds as in Sec.~\ref{sec:numerics}, but with three phases the integral over the global angle now enforces $m_1+m_2+m_3=0$, so only two \emph{signed} mode indices survive. Storing the pair $(m,n)$ for the $\Delta$- and $b$-angle modes fixes the $r'$ order to $-(m{+}n)$, and with $\v{k}$ chosen along the $x$ axis,
\begin{equation}
\begin{split}
&C_{mn}(\Delta,r',b)
=\frac{1}{(2\pi)^2}\int_0^{2\pi}\!\!\int_0^{2\pi}
e^{+\ii(m\alpha+n\beta)}\,\mathcal G\;\dd\alpha\,\dd\beta,\\
&A_{mn}(k,p,q)
=\int_0^\infty\!\!\int_0^\infty\!\!\int_0^\infty
\Delta\,r'\,b\;
J_m(k\Delta)\,J_{-(m+n)}(pr')\,J_n(qb)\;
C_{mn}\;
\dd\Delta\,\dd r'\,\dd b,\\
&\mathcal F
=(2\pi)^3\sum_{m,n}
e^{-\ii\left[-(m+n)\varphi_p+n\varphi_q\right]}\,A_{mn}(k,p,q),
\end{split}
\label{eq:threekchain}
\end{equation}
where $\varphi_p$, $\varphi_q$ are the momentum azimuths. All three phases carry $+\ii$, and the powers of $\ii$ from the three Bessel expansions cancel identically in this combination.

\subsection{Streaming the three-transform contraction}
\label{sec:threek-stream}
The coefficient $C_{mn}(\Delta,r',b)$ is too large to store on the full coordinate grid. Instead, $\mathcal G$ is evaluated for a subset of the $\Delta$ grid and successive $b$ values; for each $b$, the angular dependence is projected onto $(m,n)$ and the $b$ sum is performed immediately. The $r'$ and $\Delta$ sums are then applied in sequence,
\begin{equation}
\begin{split}
H_{mn}(q,\Delta,r')
&=\sum_{s} w^{(b),s}_{n}(q)\,C_{mn}(\Delta,r',b_s),\\
K_{mn}(q,p;\Delta)
&=\sum_{i} w^{(r'),i}_{-(m+n)}(p)\,H_{mn}(q,\Delta,r'_i),\\
A_{mn}(k,p,q)
&=\sum_{l} w^{(\Delta),l}_{m}(k)\,K_{mn}(q,p,\Delta_l).
\end{split}
\label{eq:threekstream}
\end{equation}
So the full coefficient $C_{mn}(\Delta,r',b)$ is not stored. If $H$ is too large for the full $q$ grid, the same calculation is performed for successive groups of $q$ values. For a real integrand, $A_{-m,-n}=A^{*}_{mn}$, so only $m\ge0$ is computed and the remaining modes are reconstructed by this symmetry. Algorithm~\ref{alg:threek} summarizes the calculation.

\begin{algorithm}[H]
\caption{The fully fused three-momentum calculation.}
\label{alg:threek}
\begin{algorithmic}[1]
\State build the three Filon weight families
$w^{(b)}_{n}(q)$, $w^{(r')}_{-(m+n)}(p)$, and $w^{(\Delta)}_{m}(k)$
\ForAll{successive groups of $q$ values}
  \ForAll{successive subsets of the $\Delta$ grid}
    \State set $H=0$
    \For{each $b$ value}
      \State evaluate $\mathcal G(\Delta,r',b,\alpha,\beta)$
      \State project onto $(m,n)$ and add the $b$-weighted result to $H$
    \EndFor
    \State apply the $r'$ weights to $H$ to obtain $K$
    \State apply the $\Delta$ weights to $K$ and add the result to $A$
  \EndFor
\EndFor
\end{algorithmic}
\end{algorithm}

\subsection{Angular projection at each coordinate point}
\label{sec:threek-proj}
At each $(\Delta,r',b)$ point, $\mathcal G$ is sampled on the $N_\phi\times N_\phi$ angular grid. The coefficients $C_{mn}$ are obtained by first summing over $\alpha$ and then over $\beta$, retaining only the required values of $(m,n)$. Each coefficient is multiplied by its $b$ weight and added directly to $H$ in Eq.~\eqref{eq:threekstream}; the $r'$ and $\Delta$ weights are then applied to obtain $A_{mn}$. This is the calculation stated in Algorithm~\ref{alg:threek}.

\subsection{Validation against an analytic Gaussian family}
\label{sec:threek-gates}
\begin{figure}[H]
\centering
\includegraphics[width=0.62\textwidth]{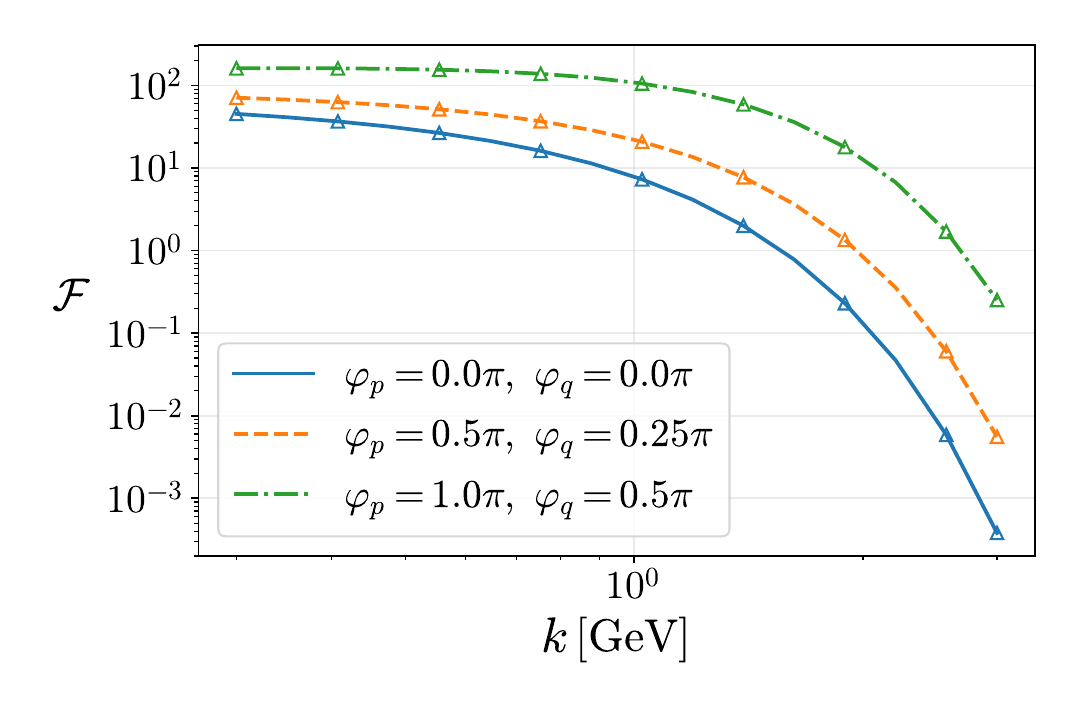}
\caption{Analytical (lines) and our numerical (open triangle points) results of the Eq.~(\ref{eq:threekchain}) for the test function in Eq.~\eqref{eq:gaussfamily}, shown as a function of $k$ at $p=0.8$ GeV and $q=1.2$ GeV for three angle pairs.}
\label{fig:threekcheck}
\end{figure}

We test the algorithm with the integrable function
\begin{equation}
\begin{split}
\mathcal G_{\rm test}
&=\exp\!\Big(
-a\Delta^2-b_0 r'^2-c\,b^2
+g_1\,\Delta r'\cos\alpha
\\
&\hspace{2.5cm}
+g_2\,r' b\cos\beta
+g_3\,\Delta b\cos(\alpha-\beta)
\Big),
\end{split}
\label{eq:gaussfamily}
\end{equation}
where the constants are chosen so that the quadratic form is positive definite, and whose six-dimensional Fourier transform is known analytically. We use $M_{\max}=25$, $N_\phi=64$, and $320$ nodes on $[0,10]~\mathrm{GeV}^{-1}$ for each radial coordinate. The numerical and exact results agree to better than $4\times10^{-6}$ at every point in Fig.~\ref{fig:threekcheck}, verifying the angular projection, the three Hankel transforms, and the final sum over angular modes.

\subsection{Outlook for physical applications}
\label{sec:threek-application}

We have not yet applied the three-transform algorithm to a specific physical process. The correlators and kinematic factors depend on the chosen process and determine both the coordinate grids and the GPU calculation. We leave this application in the next-leading order p-A CGC calculations and its performance study to future work.

\section{Conclusion}
\label{sec:conclusion}
We have developed a Filon-based algorithm for the nested two-dimensional Fourier transforms that arise from multi-Wilson-line correlators in Color Glass Condensate calculations, and validated it through an incremental sequence of implementations. Starting from the deep inelastic scattering dijet cross section without the correlation-limit approximation, we showed that the exprel identity of Eq.~\eqref{eq:exprel2} removes the numerical instability of the quadrupole correlator near the codimension-one surface where the naive parametrization takes the form $0/0$, and that a Filon-type quadrature converts each oscillatory Hankel transform into a precomputed weight vector, turning the full transform chain into a sequence of matrix products. Each mathematical reduction and each numerical step was validated before being carried into the next implementation. The CPU reference implementation of Sec.~\ref{sec:cpu} was checked against direct convergence tests in the angular mode number and the Filon grid, the CUDA table builder of Sec.~\ref{sec:tablecuda} was checked against the CPU reference, and the fused GPU implementation of Sec.~\ref{sec:fusedcuda} was checked against the table-based path. The resulting calculation evaluates the DIS dijet cross section over the full kinematic range relevant for phenomenology within a practical runtime, reducing the computation from several hours on a multi-core CPU to on the order of one to two minutes on a single commercial GPU.

The experience gained from this staged development, encapsulated in Algorithms~\ref{alg:cpu}--\ref{alg:threek}, allowed us to generalize the method to three sequential two-dimensional Fourier transforms. Because the corresponding momentum-free intermediate would be too large to store for a realistic range of Bjorken-$x$ and $\eps$ values, this general algorithm is implemented directly in the fused, streaming form of Sec.~\ref{sec:fusedcuda}. We validated the resulting six-dimensional transform against an analytic Gaussian integrand family with a closed-form solution at every stage of the calculation, confirming the angular projection, the three Hankel transforms, and the final reconstruction of the physical result.

The three-transform algorithm has not yet been deployed for a production CGC observable. Different processes involve different combinations of correlators and kinematic prefactors, so each physical application requires its own CUDA kernels, memory layout, and optimization, and a timing measurement obtained with the analytic test function would not be representative of a process-specific calculation. The production coordinate and momentum grids likewise depend on the observable under consideration. We therefore defer the production implementation and performance study to future work on next-to-leading-order proton-nucleus and electron-ion scattering cross sections computed without the back-to-back correlation-limit approximation. The present algorithm and its validated GPU implementation provide a direct foundation for these applications.

\textbf{\textit{Published Code}:} The CPU, CUDA, and fused GPU implementations described in this paper, together with the scripts used to generate the validation figures, are publicly available at \url{[https://github.com/CCNU-CGC-py/FFT_filon]}.

\section*{Acknowledgements}

{\setlength{\emergencystretch}{3em}
We thank Xiaolong Liu for reviewing the manuscript.  This work was supported in part by the Outstanding Leading Talent Team Program of Central China Normal University (XJ2026000302). This work was also supported in part by the 2025 Cross Research Project of the Fundamental Research Funds for the Central Universities of Central China Normal University, ``Advanced Detection and Artificial Intelligence at the Frontiers of Physics'' (No.~30101250317).
C. Y is supported in part by National Natural Science Foundation of China (NSFC) under Grant No. 12547164 and the China Postdoctoral Science Foundation under Grant Number 2025M783388. H. D is partially supported by NSFC under Grant No. 12535010.
}

\bibliographystyle{elsarticle-num}
\biboptions{sort&compress}
\bibliography{draft}

\end{document}